----------------------------------------------------------------

\documentstyle[aps,12pt]{revtex}
\begin{document}

\title{Independent Eigenstates of Angular Momentum\\ 
in a Quantum $N$-body System}

\author{Xiao-Yan Gu, Bin Duan and Zhong-Qi Ma \thanks{Electronic address:
MAZQ@SUN.IHEP.AC.CN}}

\address{China Center of Advanced Science and Technology (World Laboratory),\\ 
P.O.Box 8730, Beijing 100080, and \\
Institute of High Energy Physics, Beijing 100039,
The People's Republic of China}


\maketitle

\vspace{5mm}
\begin{abstract}
The global rotational degrees of freedom in the Schr\"{o}dinger 
equation for an $N$-body system are completely separated 
from the internal ones. After removing the motion of  
center of mass, we find a complete set of $(2\ell+1)$
independent base functions with the angular momentum 
$\ell$. These are homogeneous polynomials in the 
components of the coordinate vectors and the solutions of the
Laplace equation, where the Euler angles do not appear explicitly. 
Any function with given angular momentum 
and given parity in the system can be expanded with respect 
to the base functions, where the coefficients are the functions 
of the internal variables. With the right choice of the 
base functions and the internal variables, we explicitly establish 
the equations for those functions. Only $(3N-6)$ internal 
variables are involved both in the functions and in the 
equations. The permutation symmetry of the wave functions for 
identical particles is discussed. 

\vspace{3mm}
\noindent
PACS number(s): 03.65.Ge, 11.30.Er, and 03.65.Fd

\end{abstract}

\vspace{6mm}
\section{INTRODUCTION}

For a quantum $N$-body system with a pair potential, the 
Schr\"{o}dinger equation is invariant under spatial 
translation, rotation, and inversion. It is well known that,
due to the translation symmetry of the system, the wave
function can be separated into a product of two parts. One 
describes the motion of the center of mass as a free particle, 
and the other describes the motion of the system in the 
center-of-mass frame. It is no loss of generality to suppose
the center of mass of the system to be at rest, so that
the configuration is completely specified by $(N-1)$ vectors
${\bf r}_{cj}$, $1\leq j\leq N-1$, which are usually chosen as 
the Jacobi coordinate vectors ${\bf R}_{j}$ for simplicity 
\cite{eck,ohr,ma} (see Sec. II). On the other hand, due to the 
symmetries of the global rotation and space inversion of the 
system, the three rotational degrees of freedom should be separated 
completely from the internal ones so that only $(3N-6)$ internal 
variables, called the shape coordinates in some papers, are 
involved both in the functions and in the equations. This is the 
aim of this paper.

The hydrogen atom is a typical quantum two-body system, 
where there is only one Jacobi coordinate vector, which is 
proportional to the relative position vector 
${\bf r}={\bf r}_{1}-{\bf r}_{2}$. The Schr\"{o}dinger
equation for the hydrogen atom becomes a partial differential equation 
with respect to three components of ${\bf r}$. Because of the 
spherical symmetry, the angular momentum is conserved, and the 
wave function can be expressed as a product of a radial function 
$\phi(r)$ and a spherical harmonic function $Y^{\ell}_{m}(\theta,\varphi)$,
$$\Psi^{\ell}_{m}({\bf r})=\phi(r)Y^{\ell}_{m}(\theta,\varphi), \eqno (1) $$

\noindent 
where the radial function $\phi(r)$ satisfies the radial equation
containing only one radial variable. The generalization of this 
method to a quantum $N$-body system is an important and
fundamental problem that has been attacked by many groups.

Wigner studied this problem using group theory \cite{wig}. The 
global rotation of a system can be described by a space rotation 
$R=R(\alpha,\beta,\gamma)$, rotating the center-of-mass frame 
into the body-fixed frame, where $\alpha$, $\beta$ and $\gamma$ 
are the Euler angles. Briefly denoting all the internal variables 
for simplicity by $\xi$, which is invariant in the global rotation, 
one may express the wave function with a given angular momentum as 
$\Psi^{\ell}_{m}(R,\xi)=\Psi^{\ell}_{m}(\alpha,\beta,\gamma,\xi)$. 
Let $P_{S}$ be the transformation operator for a scalar
function $\psi(x)$ in the transformation $S$, 
$P_{S}\psi(x)=\psi(S^{-1}x)$ (see p. 105 in Ref. \cite{wig}). In a 
rotation $S\in$ SO(3), the function $\Psi^{\ell}_{m}(R,\xi)$ transforms as
$$P_{S}\Psi^{\ell}_{m}(R,\xi)=\Psi^{\ell}_{m}(S^{-1}R,\xi)=
\displaystyle \sum_{m'=-\ell}^{\ell}~\Psi^{\ell}_{m'}(R,\xi)
D^{\ell}_{m'm}(S). $$

\noindent
$\Psi^{\ell}_{m}(R,\xi)$ was called by Wigner \cite{wig} the function 
belonging to the $m$th row of the representation $D^{\ell}$(SO(3)). 
Letting $R=R(0,0,0)$ be the identity element, one obtains
$$\Psi^{\ell}_{m}(S^{-1},\xi)=P_{S}\Psi^{\ell}_{m}(0,0,0,\xi)
=\displaystyle \sum_{m'=-\ell}^{\ell}~\phi_{m'}^{\ell}(\xi)
D^{\ell}_{m' m}(S). $$

\noindent 
where $\phi_{m'}^{\ell}(\xi)=\Psi^{\ell}_{m'}(0,0,0,\xi)$ depends 
only on the internal variables, called the generalized radial 
functions in this paper. Due to the spherical symmetry, one only 
needs to study the eigenfunctions of angular momentum with the 
largest eigenvalue of $L_{z}$ ($m=\ell$), which in this paper
are simply called the wave functions with angular momentum $\ell$ 
for simplicity. Their partners with the smaller 
eigenvalues of $L_{z}$ can be calculated from them by the lowering 
operator $L_{-}$. Letting $S^{-1}=R(\alpha,\beta,\gamma)$, one 
obtains (see Eq. (19.6) in \cite{wig})
$$\Psi^{\ell}_{\ell}(\alpha,\beta,\gamma,\xi)
=\Psi^{\ell}_{\ell}(R,\xi)=
\displaystyle \sum_{m=-\ell}^{\ell}~D^{\ell}_{\ell m}(\alpha,\beta,\gamma)^{*}
\phi^{\ell}_{m}(\xi). \eqno (2) $$

\noindent 
where the commonly used form of the $D$function \cite{edm} is adopted. 
In Eq. (2) $D^{\ell}_{\ell m}(\alpha,\beta,\gamma)^{*}$ plays 
the role of the base function. What Wigner proved is that $(2\ell+1)$ 
functions $D^{\ell}_{\ell m}(\alpha,\beta,\gamma)^{*}$ constitute a 
complete set of independent base functions with the angular momentum 
${\ell}$, and any wave function with the angular momentum 
${\ell}$ can be expanded with respect to those base functions. 
Due to the singularity of the Euler angles, the generalized radial 
equations satisfied by the generalized radial functions are quite 
difficult to derive based on Eq. (2). Hirschfelder and Wigner \cite{hir}
studied the problem of the generalized radial equations. Later, 
the generalized radial equations were improved by several authors
\cite{cur1,cur2,pac}. The equations seem quite cumbersome \cite{cur1}.
In the present paper we are going to rechoose the base functions
as the homogeneous polynomials in the components of the coordinate 
vectors so that the derivation of the generalized radial equations 
becomes very simple. 

The generalized radial equations for a quantum three-body 
system have been discussed in more detail \cite{haf,bar1,bar2}. 
Recently, by making use of the body-fixed frame, the 
expression for the kinetic energy operator was built in terms of 
the partial angular momentum operators and radial derivatives 
containing $(3N-6)$ internal variables \cite{gat,mla1}. 
A coupled angular momentum basis was used to prediagonalize 
the kinetic energy operator, where some off-diagonal elements 
remain nonvanishing. Those results have been generalized to 
nonorthogonal vectors \cite{mla2,iun}. In those calculations,
a function with a given angular momentum was obtained from the
partial angular momentum states using Clebsch-Gordan coefficients. 
Since the partial angular momenta are generally not conserved,
one has to deal with, in principle, an infinite number of
partial angular momentum states. This problem also occurs in the
hyperspherical harmonic function method and its improved versions
\cite{viv,kri1,fan,lin,kri2,tang}. It causes unnecessary degeneracy 
of the hyperspherical harmonic states because, as Wigner proved, 
only $(2\ell+1)$ base functions with angular momentum 
$\ell$ are involved in the calculation. 

Eckart \cite{eck} presented another method, called the principal axis 
transformation, to distinguish the global rotation and the 
internal motion in a classical $N$-body system. From the 
$(N-1)$ Jacobi coordinate vector ${\bf R}_{j}$, he defined
an $(N-1) \times (N-1)$ real symmetric matrix
${\cal R}_{jk}={\bf R}_{j}\cdot {\bf R}_{k}$, which is
semipositive definite. ${\cal R}$ can be diagonalized by a 
real orthogonal similarity transformation $X$, $X{\cal R}X^{-1}=\Gamma$. 
$$\left(\displaystyle \sum_{s=1}^{N-1}~X_{js}{\bf R}_{s}\right) 
\cdot \left(\displaystyle \sum_{t=1}^{N-1}~X_{kt}{\bf R}_{t}\right) 
=\delta_{jk}\Gamma_{jj}. \eqno (3) $$

\noindent
Since there are at most three orthogonal vectors in a three 
dimensional space, the vectors can be expressed as
$$\displaystyle \sum_{k=1}^{N-1}X_{jk}{\bf R}_{k}=
\left\{\begin{array}{ll}
{\bf e}_{a}r_{a}~~~&{\rm when}~~j= a \leq 3 \\
0 &{\rm when}~~4 \leq j \leq N-1 ,\end{array} \right. \eqno (4) $$

\noindent
where ${\bf e}_{a}$ are three orthonormal vectors in the usual
three dimensional space and $r_{a}^{2}=\Gamma_{aa}$. Three 
orthonormal vectors ${\bf e}_{a}$ contain three Euler angles
describing the global rotation of the system \cite{eck}. Thus, from 
Eqs. (3) and (4), Eckart obtained 
$${\bf R}_{j}=\displaystyle \sum_{a=1}^{3}~
{\bf e}_{a}r_{a}X_{a j},~~~~~~1\leq j \leq N-1. \eqno (5) $$

\noindent
$X_{a j}$ are the first three rows of the matrix $X$ and contain 
$(3N-9)$ independent parameters. Therefore, $r_{a}X_{a j}$ contain 
$(3N-6)$ internal variables, and the rotational variables (Euler 
angles) can be completely separated in the kinetic energy expression 
from the internal variables \cite{eck}. This approach has been 
further studied and quantized in recent years \cite{ohr,cos,cha1,cha2}. 
The internal coordinates and their conjugate momenta were quantized to 
derive the kinetic energy expression through generalized angular 
momentum operators. However, the formal formula (5) does not give
the explicit functional relation of the internal coordinates  
with the components of the Jacobi coordinate vectors, so that the 
kinetic energy expression cannot be transformed directly 
from the usual kinetic energy term in the Schr\"{o}dinger equation 
by replacement of variables. It is very difficult to obtain the
wave function on the position vectors ${\bf r}_{k}$ (or on the Jacobi
coordinate vectors ${\bf R}_{j}$) from a solution on these internal 
coordinates. The intermediate calculations for the kinetic energy 
expression are so complicated that, as said in Ref. \cite{cha2}, the 
expression for a quantum six-body system has not been obtained 
probably due to a few mistakes in calculations.

Let us return to the hydrogen atom problem. After removing the 
motion of the center of mass, the configuration space is parametrized 
in terms of the rectangular coordinates ${\bf r}=(x,y,z)$ or the 
spherical coordinates $(r,\theta,\varphi)$, where $r$ specifies 
the internal (radial) motion and $(\theta,\varphi)$ specify the 
overall rotation. There is another way to separate the rotational 
degrees of freedom and obtain the same radial function and the 
radial equation as those derived from Eq. (1). One may avoid 
explicitly introducing the rotational angles $\theta$ and $\varphi$ 
using the harmonic polynomial
${\cal Y}^{\ell}_{m}({\bf r})=r^{\ell}Y^{\ell}_{m}(\theta,\varphi)$,
which is a homogeneous polynomial of degree $\ell$ in the
rectangular coordinates $(x,y,z)$ and satisfies the Laplace equation
as well as the eigenequation of the angular momentum. 
Using ${\cal Y}^{\ell}_{m}({\bf r})$, Eq. (1) becomes
$$\Psi^{\ell}_{m}({\bf r})=\left\{r^{-\ell}\phi(r)\right\}
{\cal Y}^{\ell}_{m}({\bf r}). $$

\noindent
Under the action of the Laplace operator, we have
$$\bigtriangleup \Psi^{\ell}_{m}({\bf r})
={\cal Y}^{\ell}_{m}({\bf r})
\left[\bigtriangleup r^{-\ell}\phi(r)\right]
+2\bigtriangledown \left[r^{-\ell}\phi(r)\right]
\cdot \bigtriangledown \left\{{\cal Y}^{\ell}_{m}({\bf r})\right\}$$
$$={\cal Y}^{\ell}_{m}({\bf r})r^{-1}\partial_{r}^{2}r
\left[r^{-\ell}\phi(r)\right]
+2\partial_{r} \left[r^{-\ell}\phi(r)\right]~r^{-1}{\bf r} \cdot
\bigtriangledown \left\{{\cal Y}^{\ell}_{m}({\bf r})\right\}$$
$$={\cal Y}^{\ell}_{m}({\bf r})r^{-\ell}
\left\{r^{-1}\partial_{r}^{2}r\phi(r)
+\ell(\ell+1) r^{-2} \phi(r)
+2 \left(-\ell r^{-1}\right)\left[\partial_{r}\phi(r)+r^{-1}\phi(r)\right]
\right\}$$
$$+2\left\{-\ell r^{-\ell-1}\phi(r)+r^{-\ell}\partial_{r}\phi(r)\right\}
\left\{\ell r^{-1}{\cal Y}^{\ell}_{m}({\bf r})\right\}$$
$$=Y^{\ell}_{m}(\theta,\varphi)\left\{r^{-1}\partial_{r}^{2}r\phi(r)
-\ell(\ell+1) r^{-2} \phi(r)\right\}, $$

\noindent
where and hereafter $\partial_{r}\psi$ denotes $\partial \psi/\partial r$
and so on. The results are the same. In the traditional approach, 
the property that $Y^{\ell}_{m}(\theta,\varphi)$ is the eigenfunction
of $L^{2}$ is used, and in the new approach, the property 
that ${\cal Y}^{\ell}_{m}({\bf r})$ is a homogeneous
polynomial in the rectangular coordinates and the solution to
the Laplace equation is used. It is worth emphasizing that the
rotational angles $\theta$ and $\varphi$ do not appear explicitly
in this approach. Since the differential calculus with
respect to $\theta$ and $\varphi$ is not complicated, 
this approach is similar to the traditional one in a two-body system.
However, it may be easier in an
$N$-body system due to the complicated calculus with respect to
the Euler angles. In the present paper we will separate the 
global rotational variables in the Schr\"{o}dinger equation 
for an $N$-body system from the internal ones by a generalized 
method following the approach described above. In our approach, the number
of base functions with the given angular momentum is finite, 
but that number in the hyperspherical harmonic function method
and its improved versions \cite{gat,viv,kri1,fan,lin,kri2,tang}
is infinite due to the unconserved partial angular momenta. 
We also avoid the heavy differential calculus with 
respect to the Euler angles which is sometimes necessary for 
expressing kinetic energy operators. 

This paper is organized as follows. In Sec. II we will 
briefly review the method of separating the motion of center of mass 
by the Jacobi coordinate vectors. In Sec. III we will define 
$(3N-6)$ internal variables from the Jacobi coordinate vectors 
${\bf R}_{j}$ and find the $(2\ell+1)$ base functions with 
total orbital angular momentum $\ell$, which are the homogeneous 
polynomials in the components of ${\bf R}_{j}$ and the solutions 
of the Laplace equation. Then we will prove that the base functions 
constitute a complete set, namely, any function with the angular 
momentum $\ell$ and the given parity in the system can be expanded 
with respect to the base functions, where the coefficients depend 
only on the internal variables. Since the base functions are 
polynomials, we are able to derive easily the generalized 
radial equations satisfied by the coefficients explicitly in 
Sec. IV. The permutation symmetry for the total wave function 
when some or all of the particles in the system are identical particles 
is discussed in Sec. V. In Sec. VI we will derive the 
generalized radial equations in a general case where the Jacobi 
coordinate vectors ${\bf R}_{j}$ (orthogonal vectors) 
are replaced by arbitrary coordinate vectors ${\bf r}_{cj}$ in the 
center-of-mass frame (nonorthogonal vectors). In Sec. VII we will 
discuss a physical application of our approach. Some 
conclusions are given in Sec. VIII.

\section{Separation of Motion of Center-of-Mass}

For a quantum $N$-body system, we denote the position vectors and the 
masses of $N$ particles by ${\bf r}_{k}$ and by $m_{k}$, 
$k=1,2,\ldots,N$, respectively. $M=\sum_{k} m_{k}$ is the total 
mass. The Schr\"{o}dinger equation for the $N$-body system is 

$$- \displaystyle {\hbar^{2} \over 2} \displaystyle 
\sum_{k=1}^{N}~\displaystyle m_{k}^{-1} \bigtriangleup_{{\bf r}_{k}} \Psi 
+V \Psi =E \Psi , \eqno (6) $$

\noindent
where $\bigtriangleup_{{\bf r}_{k}}$ is the Laplace operator with 
respect to the position vector ${\bf r}_{k}$, and $V$ is a pair 
potential, depending upon the distance of each pair of particles, 
$|{\bf r}_{j}-{\bf r}_{k}|$.

Now, we replace the position vectors ${\bf r}_{k}$ by the Jacobi
coordinate vectors ${\bf R}_{j}$:
$${\bf R}_{0}=M^{-1/2}\displaystyle \sum_{k=1}^{N}~m_{k}{\bf r}_{k},~~~~~
{\bf R}_{j}=\left(\displaystyle {m_{j}M_{j+1}\over M_{j}}
 \right)^{1/2}\left({\bf r}_{j}-\displaystyle \sum_{k=j+1}^{N}~
\displaystyle {m_{k}{\bf r}_{k}\over M_{j+1}}\right), $$
$$1\leq j \leq (N-1),~~~~~~M_{j}=\displaystyle 
\sum_{k=j}^{N}~m_{k},~~~~~~M_{1}=M,  \eqno (7) $$

\noindent
where ${\bf R}_{0}$ describes the position of the center of mass,
${\bf R}_{1}$ describes the mass-weighted separation from the first 
particle to the center of mass of the remaining particles, ${\bf R}_{2}$ 
describes the mass-weighted separation from the second particle
to the center of mass of the remaining $N-2$ particles, and so on.
The mass-weighted factors in front of the formulas 
for ${\bf R}_{j}$ are determined by the condition
$$\sum_{k=1}^{N}~m_{k}{\bf r}_{k}^{2}
=\sum_{j=0}^{N-1}~{\bf R}_{j}^{2}, $$ 

\noindent
where an additional factor $\sqrt{M}$ is included in ${\bf R}_{j}$ 
for convenience. One may determine the factors one by one from the 
following schemes. In the center-of-mass frame, if the first 
$j-1$ particles are located at the origin and the 
last $N-j$ particles coincide with each other, the factor 
in front of ${\bf R}_{j}$ is determined by
$${\bf r}_{j+1}={\bf r}_{j+2}=\cdots ={\bf r}_{N}
=-m_{j}{\bf r}_{j}/M_{j+1},~~~~~
\displaystyle \sum_{k=j}^{N}~m_{k}{\bf r}_{k}^{2}
={\bf R}_{j}^{2}. \eqno (8) $$

A straightforward calculation by replacement of variables shows 
that the Laplace operator in Eq. (6) and the orbital angular 
momentum operator ${\bf L}$ are directly expressed in ${\bf R}_{j}$:
$$\bigtriangleup =\displaystyle \sum_{k=1}^{N}~\displaystyle
m_{k}^{-1} \bigtriangleup_{{\bf r}_{k}}
= \displaystyle \sum_{j=0}^{N-1}~
\bigtriangleup_{{\bf R}_{j}}, $$
$${\bf L}=-i\hbar \displaystyle \sum_{k=1}^{N}~ {\bf r}_{k}
\times \bigtriangledown_{{\bf r}_{k}}
=-i\hbar \displaystyle \sum_{j=0}^{N-1}~ {\bf R}_{j}
\times \bigtriangledown_{{\bf R}_{j}},   \eqno (9) $$

In the center-of-mass frame, ${\bf R}_{0}=0$. Since the Laplace 
operator does not contain mixed derivative terms, the
Jacobi coordinate vectors are also called the orthogonal
vectors \cite{mla1}. The Laplace operator 
obviously has the symmetry of the O$(3N-3)$ group with respect to 
$(3N-3)$ components of $(N-1)$ Jacobi coordinate vectors. The 
O$(3N-3)$ group contains a subgroup SO(3)$\times$O$(N-1)$, where 
SO(3) is the usual rotation group. The space inversion and 
the different definitions for the Jacobi coordinate vectors in the 
so-called Jacobi tree \cite{fan} can be obtained by O$(N-1)$ 
transformations. For the system of identical particles, the 
permutation group among particles is also a subgroup of the O$(N-1)$ 
group. As a matter of fact, after the transposition $(k,k+1)$ 
between the $k$th and the $(k+1)$th particles, the new Jacobi 
coordinate vectors, denoted by ${\bf R}^{\prime}_{j}$, satisfy
$${\bf R}^{\prime}_{j}={\bf R}_{j},~~~~~{\rm when}~~
j \neq k ~~{\rm or}~~k+1, $$
$${\bf R}^{\prime}_{k}=\left[\displaystyle {m_{k+1} \over
M_{k}\left(m_{k}+M_{k+2}\right)}\right]^{1/2}
\left[\left(m_{k}+M_{k+2}\right){\bf r}_{k+1}
-m_{k}{\bf r}_{k}-\displaystyle \sum_{j=k+2}^{N}~
m_{j}{\bf r}_{j} \right]$$
$$=-{\bf R}_{k}\cos \theta_{k}+{\bf R}_{k+1}\sin \theta_{k},$$
$${\bf R}^{\prime}_{k+1}=\left[\displaystyle {m_{k} \over
\left(m_{k}+M_{k+2}\right)M_{k+2}}\right]^{1/2}
\left[M_{k+2}{\bf r}_{k}-\displaystyle \sum_{j=k+2}^{N}~
m_{j}{\bf r}_{j} \right]$$
$$={\bf R}_{k}\sin \theta_{k}+{\bf R}_{k+1}\cos \theta_{k},$$
$$\cos \theta_{k}=\left[\displaystyle {m_{k}m_{k+1} \over
M_{k+1}\left(m_{k}+M_{k+2}\right)}\right]^{1/2},~~~~~~~~
\sin \theta_{k}=\left[\displaystyle {M_{k}M_{k+2} \over
M_{k+1}\left(m_{k}+M_{k+2}\right)}\right]^{1/2}. \eqno (10) $$

\noindent
This is obviously an O$(N-1)$ transformation. For a system of 
identical particles, $\cos\theta_{k}=(N-k)^{-1}$. 

It is easy to obtain the inverse transformation of Eq. (7):
$${\bf r}_{j}=\left[M_{j+1} \over m_{j}M_{j}\right]^{1/2}{\bf R}_{j}
-\displaystyle \sum_{k=1}^{j-1}~
\left[m_{k} \over M_{k}M_{k+1}\right]^{1/2}{\bf R}_{k}
+M^{-1/2}{\bf R}_{0}, \eqno (11) $$
$${\bf r}_{j}-{\bf r}_{k}=
\left[M_{j+1} \over m_{j}M_{j}\right]^{1/2}{\bf R}_{j}
-\displaystyle \sum_{i=k+1}^{j-1}~
\left[m_{i} \over M_{i}M_{i+1}\right]^{1/2}{\bf R}_{i}
-\left[M_{k} \over m_{k}M_{k+1}\right]^{1/2}{\bf R}_{k}. \eqno (12) $$

\noindent
Thus, the potential $V$ is a function of 
${\bf R}_{j}\cdot {\bf R}_{k}$.

The Jacobi coordinate vectors ${\bf R}_{j}$ are invariant in 
translation and constitute a complete set of the coordinate vectors 
in the center-of-mass frame. If a complete set of arbitrary coordinate
vectors ${\bf r}_{cj}$ is chosen to replace the Jacobi coordinate 
vectors,  
$${\bf r}_{cj}=\displaystyle \sum_{k=1}^{N-1}~{\bf R}_{k}D_{kj}, 
~~~~~~\det D \neq 0, \eqno (13) $$

\noindent
where $D_{kj}$ are functions of the masses $m_{j}$, the Laplace 
operator and the angular momentum operator become 
$$\bigtriangleup = \displaystyle \sum_{j,k=1}^{N-1}~S_{jk}
\bigtriangledown_{{\bf r}_{cj}} \cdot \bigtriangledown_{{\bf r}_{ck}},
~~~~~S_{jk}=\displaystyle \sum_{t=1}^{N-1}~D_{tj}D_{tk}, ~~~~~~
{\bf L}=-i\hbar \displaystyle \sum_{j=1}^{N-1}~ {\bf r}_{cj}
\times \bigtriangledown_{{\bf r}_{cj}}.   \eqno (14) $$

\noindent
A typical example is 
${\bf r}_{cj}={\bf r}_{j}-M^{-1/2}{\bf R}_{0}$ [see Eq. (11)]. 
When the matrix $S$ is not diagonal, ${\bf r}_{cj}$ are called
the nonorthogonal vectors \cite{mla2,iun}. We will not discuss
the nonorthogonal vectors until Sec. VI. 

\section{BASE FUNCTIONS WITH THE GIVEN ANGULAR MOMENTUM}

Because of the spherical symmetry, the angular momentum is
conserved. We are going to discuss the wave functions with
the given angular momentum and parity. From the given 
form (9), the eigenfunctions of the angular momentum 
${\bf L}^{2}$ are the homogeneous polynomials in the 
components $R_{jb}$ of the Jacobi coordinate vectors ${\bf R}_{j}$.

For a quantum two-body system, there is only one Jacobi coordinate 
vector ${\bf R=r}$, and the eigenfunction of the angular momentum 
is the spherical harmonic function $Y^{\ell}_{m}(\theta, \varphi)$. 
What is the generalization of the spherical harmonic function 
for a quantum $N$-body system? A naive idea for generalization 
is to introduce the Euler angles, as was done by Wigner 
\cite{wig,cur1,cur2,pac}. Is it necessary to introduce angular 
variables in the eigenfunction of the angular momentum?

As is well known, the harmonic polynomial 
${\cal Y}^{\ell}_{m}({\bf r})=r^{\ell}Y^{\ell}_{m}(\theta, \varphi)$
is a homogeneous polynomial of degree $\ell$ in the components of 
${\bf r}$, which satisfies the Laplace equation as well as the 
eigenequation of the angular momentum. It does not contain  
angular variables explicitly. The number of linearly independent 
homogeneous polynomials of degree $\ell$ in the components of 
${\bf r}$ is
$$N(\ell)= \sum_{s=0}^{\ell}(\ell-s+1)=(\ell+1)(\ell+2)/2. $$
  
\noindent
The number of homogeneous polynomials that can be expressed
as a product of ${\bf r\cdot r}$ and a homogeneous polynomial of 
degree $(\ell-2)$ is $N(\ell-2)$. Because $N(\ell)-N(\ell-2)=2\ell+1$,
the remaining homogeneous polynomials of degree $\ell$ are nothing but
the harmonic polynomials ${\cal Y}^{\ell}_{m}({\bf r})$. 

For a quantum three-body system there are two Jacobi coordinate vectors 
${\bf R}_{1}$ and ${\bf R}_{2}$ and three internal variables in 
the center-of-mass frame:  

$$\xi_{1}={\bf R}_{1}\cdot {\bf R}_{1},~~~~~
\xi_{2}={\bf R}_{1}\cdot {\bf R}_{2},~~~~~
\eta_{2}={\bf R}_{2}\cdot {\bf R}_{2}. \eqno (15) $$

\noindent
The internal variables are invariant in the global rotation 
and the space inversion of the system. We are going to 
construct base functions for angular momentum that do not 
contain a function of the internal variables as
a multiplying factor, because the factor should be incorporated 
into the generalized radial functions. The number of linearly 
independent homogeneous polynomials of degree $\ell$ in the 
components of the Jacobi coordinate vectors is $M(\ell)$: 
$$M(\ell)=\displaystyle {1\over 5!}(\ell+1)(\ell+2)(\ell+3)
(\ell+4)(\ell+5). $$

\noindent
The number of the homogeneous polynomials of degree $\ell$ that 
do not contain a function of internal variables as a factor is 
$$K(\ell)=M(\ell)-3M(\ell-2)+3M(\ell-4)-M(\ell-6)
=4\ell^{2}+2,~~~~~\ell \geq 1. $$

On the other hand, the wave function with angular momentum 
$\ell$ can be obtained from 
${\cal Y}^{q}_{m}({\bf R}_{1}){\cal Y}^{\ell-q}_{m'}({\bf R}_{2})$
by use of the Clebsch-Gordan coefficients 
$\langle q,m,\ell-q,\mu -m| L, \mu \rangle$ \cite{edm}:
$${\cal Y}^{\ell q}_{L \mu}({\bf R}_{1},{\bf R}_{2})
=\displaystyle \sum_{m}~{\cal Y}^{q}_{m}({\bf R}_{1})
{\cal Y}^{\ell-q}_{\mu-m}({\bf R}_{2})
\langle q,m,\ell-q,\mu -m| L, \mu \rangle. \eqno (16) $$

\noindent
${\cal Y}^{\ell q}_{L \mu}({\bf R}_{1},{\bf R}_{2})$ is a 
homogeneous polynomial of degree $\ell$ in the components of 
the Jacobi coordinate vectors ${\bf R}_{j}$. Simultaneously, 
it is the common eigenfunction of ${\bf L}^{2}$, $L_{z}$ and 
the space inversion with eigenvalues $L(L+1)$, $\mu$ and 
$(-1)^{\ell}$, respectively. When $\mu=L=\ell$ ($0\leq q \leq \ell$) 
and $(\ell-1)$ ($1\leq q \leq \ell-1$), we have
$${\cal Y}^{\ell q}_{\ell \ell}({\bf R}_{1},{\bf R}_{2})
=(-1)^{\ell} \left\{ \displaystyle 
{ \left[(2q+1)!(2\ell-2q+1)!\right]^{1/2} \over q! (\ell-q)! 2^{\ell+2}\pi}
\right\} (R_{1x}+i R_{1y})^{q}(R_{2x}+iR_{2y})^{\ell-q} , $$
$${\cal Y}^{\ell q}_{(\ell-1)(\ell-1)}({\bf R}_{1},{\bf R}_{2})
=(-1)^{\ell-1} \left\{ \displaystyle 
{ (2q+1)!(2\ell-2q+1)! \over 2q\ell (\ell-q)} \right\}^{1/2}
\left\{(q-1)! (\ell-q-1)! 2^{\ell+1}\pi \right\}^{-1}$$
$$~~~~~~\times~(R_{1x}+i R_{1y})^{q-1}(R_{2x}+iR_{2y})^{\ell-q-1} 
\left\{(R_{1x}+iR_{1y})R_{2z}-R_{1z}(R_{2x}+iR_{2y})\right\}.
 \eqno (17) $$

\noindent
It is evident that these expressions do not contain a function of the 
internal variables as a factor, neither do their partners with 
smaller $\mu$ due to the spherical symmetry. The number of 
those eigenfunctions is
$$(2\ell+1)(\ell+1)+(2\ell-1)(\ell-1)=4\ell^{2}+2=K(\ell),
~~~~~\ell \geq 1.  $$

\noindent
That is, any of the remaining eigenfunctions 
${\cal Y}^{\ell q}_{L \mu}({\bf R}_{1},{\bf R}_{2})$ with $L<\ell-1$ 
can be expressed as a combination, where each term is a product 
of a function of the internal variables and a homogeneous 
polynomial of degree less than $\ell$ \cite{gu}. For example, 
$${\cal Y}^{21}_{00}({\bf R}_{1},{\bf R}_{2})=
-\displaystyle {\sqrt{3} \over 4\pi} \xi_{2},~~~~~~
{\cal Y}^{42}_{00}({\bf R}_{1},{\bf R}_{2})=
\displaystyle {\sqrt{5} \over 8\pi}
\left\{3\xi_{2}^{2}-\xi_{1}\eta_{2}\right\},$$
$${\cal Y}^{42}_{22}({\bf R}_{1},{\bf R}_{2})=
\displaystyle {5\sqrt{21} \over 56\pi}\left\{ 
\eta_{2}(R_{1x}+i R_{1y})^{2}+\xi_{1}(R_{2x}+i R_{2y})^{2}\right. $$
$$~~~\left.-3\xi_{2}(R_{1x}+i R_{1y})(R_{2x}+i R_{2y})\right\}.   $$

\noindent
In other words, any eigenfunction with angular momentum 
$\ell$ is a combination of those homogeneous 
polynomials ${\cal Y}^{\ell q}_{\ell \ell}({\bf R}_{1},{\bf R}_{2})$ 
and ${\cal Y}^{(\ell+1) q}_{\ell \ell}({\bf R}_{1},{\bf R}_{2})$, 
where the combinative coefficients are functions of the 
internal variables. Since the normalization factor can be ignored, 
we rewrite
${\cal Y}^{(\ell+\lambda) q}_{\ell \ell}({\bf R}_{1},{\bf R}_{2})$
in a simplified form as
$Q_{q}^{\ell \lambda}({\bf R}_{1},{\bf R}_{2})$ 
by removing a constant factor
$$Q_{q}^{\ell \lambda}({\bf R}_{1},{\bf R}_{2})
=\displaystyle {X^{q-\lambda}Y^{\ell-q}Z^{\lambda}
\over (q-\lambda)!(\ell-q)!},~~~~~~
\lambda \leq q \leq \ell,~~~~~~\lambda=0,1 $$
$$X\equiv R_{1x}+iR_{1y},~~~~Y\equiv R_{2x}+iR_{2y},~~~~
Z\equiv XR_{2z}-R_{1z}Y.  \eqno (18) $$

\noindent
Note that 
$$Q_{q}^{\ell 1}({\bf R}_{1},{\bf R}_{2})=
Q_{q-1}^{(\ell-1) 0}({\bf R}_{1},{\bf R}_{2})Z. \eqno (19) $$

\noindent
$Q_{q}^{\ell \lambda}({\bf R}_{1},{\bf R}_{2})$, called the 
generalized harmonic polynomial, is a homogeneous polynomial 
of degree $(\ell+\lambda)$ in the components of the Jacobi 
coordinate vectors. It is the common eigenfunction of 
${\bf L}^{2}$, $L_{z}$, ${\bf L}_{{\bf R}_{1}}^{2}$, 
${\bf L}_{{\bf R}_{2}}^{2}$, $\bigtriangleup_{{\bf R}_{1}}$, 
$\bigtriangleup_{{\bf R}_{2}}$, 
$\bigtriangledown_{{\bf R}_{1}}\cdot \bigtriangledown_{{\bf R}_{2}}$, 
and the space inversion 
with the eigenvalues $\ell(\ell+1)$, $\ell$, $q(q+1)$, 
$(\ell-q+\lambda)(\ell-q+\lambda+1)$, $0$, $0$, $0$, and 
$(-1)^{\ell+\lambda}$, respectively, 
where ${\bf L}_{{\bf R}_{1}}^{2}$ (${\bf L}_{{\bf R}_{2}}^{2}$) 
is the square of the partial angular momentum, and 
$\bigtriangleup_{{\bf R}_{1}}$ ($\bigtriangleup_{{\bf R}_{2}}$)
is the Laplace operator with respect to the Jacobi coordinate 
vector ${\bf R}_{1}$ (${\bf R}_{2}$) [see Eq. (9)]. Any wave 
function with the given angular momentum $\ell$ 
and the parity $(-1)^{\ell+\lambda}$ can be expressed as follows:
$$\Psi^{\ell \lambda}_{\ell}({\bf R}_{1},{\bf R}_{2})
=\displaystyle \sum_{q=\lambda}^{\ell}~
\psi^{\ell \lambda}_{q}(\xi_{1},\xi_{2},\eta_{2})
Q_{q}^{\ell \lambda}({\bf R}_{1},{\bf R}_{2}),
~~~~~\lambda=0,1. \eqno (20) $$

\noindent
That is, for a three-body system the generalized harmonic 
polynomials $Q_{q}^{\ell \lambda}({\bf R}_{1},{\bf R}_{2})$ constitute
a complete set of base functions with angular momentum
$\ell$ and parity $(-1)^{\ell+\lambda}$. 
Only $\ell+1-\lambda$ partial angular momentum states are 
involved in constructing a function with angular momentum 
$\ell$ and parity $(-1)^{\ell+\lambda}$, and 
the contributions from the infinite number of remaining 
partial angular momentum states are incorporated into those of 
the radial functions. Substituting Eq. (20) into the 
Schr\"{o}dinger equations (6) and (9), one is able to easily 
derive the generalized radial equations for the generalized 
radial functions $\psi^{\ell \lambda}_{q}(\xi_{1},\xi_{2},\eta_{2})$ 
\cite{ma,hsi}:
$$\bigtriangleup \psi^{\ell \lambda}_{q} +4q \partial_{\xi_{1}} 
\psi^{\ell \lambda}_{q} +4(\ell-q+\lambda) \partial_{\eta_{2}} 
\psi^{\ell \lambda}_{q} +2(q-\lambda) \partial_{\xi_{2}} 
\psi^{\ell \lambda}_{q-1} +2(\ell-q) \partial_{\xi_{2}} 
\psi^{\ell \lambda}_{q+1} $$
$$~~~~~~=-\displaystyle {2\over \hbar^{2}}
\left(E-V\right) \psi^{\ell \lambda}_{q},$$
$$\bigtriangleup \psi^{\ell \lambda}_{q}(\xi_{1},\xi_{2},\eta_{2})=
\left\{ 4\xi_{1}\partial^{2}_{\xi_{1}}+4\eta_{2}\partial^{2}_{\eta_{2}}
+6\left(\partial_{\xi_{1}}+\partial_{\eta_{2}}\right)\right. $$
$$\left.~~~~~~~~~+\left(\xi_{1}+\eta_{2}\right)\partial^{2}_{\xi_{2}}
+4\xi_{2}\left(\partial_{\xi_{1}}+\partial_{\eta_{2}}\right)
\partial_{\xi_{2}}\right\}\psi^{\ell \lambda}_{q}(\xi_{1},\xi_{2},\eta_{2}),$$
$$~~~~~~~~~~~~~~\lambda\leq q \leq \ell,~~~~~~~~~\lambda =0,1.   \eqno (21) $$

For a quantum $N$-body system, there are $(N-1)$ Jacobi coordinate
vectors. We arbitrarily choose two Jacobi coordinate vectors, say 
${\bf R}_{1}$ and ${\bf R}_{2}$. We fix the body-fixed frame
such that ${\bf R}_{1}$ is parallel with its $Z$-axis, and 
${\bf R}_{2}$ is located in its $XZ$ plane with a non-negative 
$x$-component. Define $(3N-6)$ internal variables, which 
are invariant in the global rotation of the system: 
$$\xi_{j}={\bf R}_{j}\cdot {\bf R}_{1},~~~~~
\eta_{j}={\bf R}_{j}\cdot {\bf R}_{2},~~~~~
\zeta_{j}={\bf R}_{j}\cdot \left({\bf R}_{1} \times 
{\bf R}_{2} \right),$$
$$1\leq j \leq (N-1),~~~~~~\eta_{1}=\xi_{2},~~~~~~
\zeta_{1}=\zeta_{2}=0.  \eqno (22) $$

\noindent 
It is worth mentioning that $\xi_{j}$ and $\eta_{j}$ have even 
parity, but $\zeta_{j}$ has odd parity. From them we have
$$\Omega_{j}=\left({\bf R}_{1} \times {\bf R}_{j}\right)\cdot
\left({\bf R}_{1}\times {\bf R}_{2}\right)
=\xi_{1}\eta_{j}-\xi_{2}\xi_{j},$$ 
$$\omega_{j}=\left({\bf R}_{2}
\times {\bf R}_{j}\right)\cdot \left({\bf R}_{1}\times {\bf
R}_{2}\right) =\xi_{2}\eta_{j}-\eta_{2}\xi_{j},$$
$$\Omega_{1}=\omega_{2}=0,~~~~~~
\Omega_{2}=-\omega_{1}=\left({\bf R}_{1}\times {\bf R}_{2}\right)^{2}.
 \eqno (23) $$

Due to our choice of the body-fixed frame, the components of 
${\bf R}_{1}$ and ${\bf R}_{2}$ in the frame are 
$\left(0,0,\xi_{1}^{1/2}\right)$ and $\left[
\left(\Omega_{2}/\xi_{1}\right)^{1/2},0,\xi_{2}\xi_{1}^{-1/2}
\right]$, respectively. From Eq. (22) we are able to express all 
the components ${\bf R}^{\prime}_{j b}$ of the Jacobi coordinate 
vectors ${\bf R}_{j}$ in the body-fixed frame by the internal 
variables: 
$$R^{\prime}_{j x}=\Omega_{j}\left(\xi_{1}\Omega_{2}
\right)^{-1/2},~~~~~R^{\prime}_{j y}=\zeta_{j}\Omega_{2}^{-1/2},
~~~~~R^{\prime}_{j z}=\xi_{j}\xi_{1}^{-1/2}. \eqno (24) $$

\noindent
The formulas (24) also hold for $j=1$ and $2$. The volume 
element of the configuration space can be calculated from
the Jacobi determinant by replacement of variables:
$$\displaystyle \prod_{j=1}^{N-1}~dR_{jx}dR_{jy}dR_{jz}=\displaystyle 
{1 \over 4}\Omega_{2}^{3-N}\sin\beta d\alpha d\beta d\gamma
d\xi_{1}d\xi_{2}d\eta_{2}\displaystyle \prod_{j=3}^{N-1}~
d \xi_{j}d\eta_{j}d\zeta_{j}. \eqno (25) $$

\noindent
The ranges of definition of the Euler angles are well known, the 
ranges of definition of $\xi_{1}$ and $\eta_{2}$ are $(0,\infty)$
and the ranges of definition of the remaining variables are 
$(-\infty,\infty)$.

Furthermore,
$${\bf R}_{j}\cdot {\bf R}_{k}=\Omega_{2}^{-1}
\left(\Omega_{j}\eta_{k}-\omega_{j}\xi_{k}
+\zeta_{j}\zeta_{k}\right). \eqno (26) $$

\noindent
It is easy to see from Eqs. (12) and (26) that the potential $V$
is a function of only the internal variables. Since ${\bf R}_{1}$ 
and ${\bf R}_{2}$ determine the body-fixed frame completely, it
also can be seen from Eq. (24) that each of the components of 
the Jacobi coordinate vectors ${\bf R}_{j}$ can be expressed 
as a linear combination of ${\bf R}_{1b}$ and ${\bf R}_{2b}$
with the coefficients depending on the internal variables. In fact, 
denote the rotation transforming the center-of-mass frame to 
the body-fixed frame by $R(\alpha, \beta, \gamma)$ with three 
Euler angles \cite{edm}
$$R(\alpha, \beta, \gamma)=\left(\begin{array}{ccc}
c_{\alpha}c_{\beta}c_{\gamma}-s_{\alpha}s_{\gamma}
&~~-c_{\alpha}c_{\beta}s_{\gamma}-s_{\alpha}c_{\gamma}~~
&c_{\alpha}s_{\beta} \\
s_{\alpha}c_{\beta}c_{\gamma}+c_{\alpha}s_{\gamma}
&-s_{\alpha}c_{\beta}s_{\gamma}+c_{\alpha}c_{\gamma}
&s_{\alpha}s_{\beta} \\
-s_{\beta}c_{\gamma}&s_{\beta}s_{\gamma}&c_{\beta} \\
\end{array} \right), \eqno (27) $$

\noindent
where $c_{\alpha}=\cos \alpha$, $s_{\alpha}=\sin \alpha$, and so on.
It is straightforward to obtain from Eqs. (18), (24) and (27) that
$$X=R_{1x}+iR_{1y}=\xi_{1}^{1/2}e^{i\alpha}s_{\beta}, $$
$$Y=R_{2x}+iR_{2y}=\left(\Omega_{2}/\xi_{1}\right)^{1/2}e^{i\alpha}
\left(c_{\beta}c_{\gamma}+is_{\gamma}\right)
+\xi_{2}\xi_{1}^{-1/2}e^{i\alpha} s_{\beta}, $$
$$R_{1z}=\xi_{1}^{1/2}c_{\beta},~~~~~~
R_{2z}=-\left(\Omega_{2}/\xi_{1}\right)^{1/2}s_{\beta}c_{\gamma}
+\xi_{2}\xi_{1}^{-1/2}c_{\beta}, $$
$$Z=\left(R_{1x}+iR_{1y}\right)R_{2z}
-R_{1z}\left(R_{2x}+iR_{2y}\right)=-\Omega_{2}^{1/2}e^{i\alpha}
\left(c_{\gamma}+ic_{\beta}s_{\gamma}\right).  \eqno (28) $$
$$R_{j x}+iR_{j y}=\Omega_{2}^{-1}\left\{
-\omega_{j}X+\Omega_{j}Y-i\zeta_{j}Z\right\}, $$
$$\left(R_{j x}+iR_{j y}\right)R_{k z}
-R_{j z}\left(R_{k x}+iR_{k y}\right)
=\Omega_{2}^{-1}\left\{i\left(\eta_{j}\zeta_{k}
-\eta_{k}\zeta_{j}\right)X\right.$$
$$~~~~~~~~\left. -i\left(\xi_{j}\zeta_{k}-\xi_{k}\zeta_{j}\right)Y
+\left(\xi_{j}\eta_{k}-\xi_{k}\eta_{j}\right)Z\right\}.  \eqno (29) $$

\noindent
Therefore, each harmonic polynomial 
${\cal Y}^{\ell}_{\ell}({\bf R}_{j})$ can be expressed as
a combination of $Q_{q}^{\ell \lambda}({\bf R}_{1},{\bf R}_{2})$
with the coefficients depending on the internal variables. This
means that the generalized harmonic polynomials 
$Q_{q}^{\ell \lambda}({\bf R}_{1},{\bf R}_{2})$ 
given in Eq. (18) do constitute a complete set of independent 
base functions with the given angular momentum 
$\ell$ for a quantum $N$-body system, just like they do for
a quantum three-body system.

Because this conclusion plays a key role in separating the 
global rotational degrees of freedom from the internal ones
in the quantum $N$-body system, we are going to prove it 
by another method. From Eq. (28) we have
$$e^{i\alpha}s_{\beta}=\xi_{1}^{-1/2}X,~~~~~~
e^{i\alpha}\left(c_{\gamma}+ic_{\beta}s_{\gamma}\right)=
-\Omega_{2}^{-1/2}Z,$$
$$e^{i\alpha}\left(c_{\beta}c_{\gamma}+is_{\gamma}\right)
=-\xi_{2}\left(\xi_{1}\Omega_{2}\right)^{-1/2}X
+\left(\xi_{1}/\Omega_{2}\right)^{1/2}Y,  \eqno (30) $$
$$Z^{2}=\eta_{2}X^{2}-2\xi_{2}XY+\xi_{1}Y^{2}. 
 \eqno (31) $$

\noindent
That is, a homogeneous polynomial of degree $\ell$ in three variables
$e^{i\alpha}s_{\beta}$, 
$e^{i\alpha}\left(c_{\beta}c_{\gamma}+is_{\gamma}\right)$, and
$e^{i\alpha}\left(c_{\gamma}+ic_{\beta}s_{\gamma}\right)$ can be
expanded with respect to $Q_{q}^{\ell \tau}({\bf R}_{1},{\bf R}_{2})$
where the coefficients only depend on the internal variables $\xi_{j}$,
$\eta_{j}$, and $\zeta_{j}$. On the other hand, the Wigner 
$D$function is \cite{edm}
$$D^{\ell}_{\ell (\pm m)}(\alpha,\beta,\gamma)^{*}=
(-1)^{\ell-m}2^{-\ell}\left[\displaystyle {(2\ell)! \over
(\ell+m)!(\ell-m)!}\right]^{1/2}e^{i(\ell \alpha\pm m\gamma)}
s_{\beta}^{\ell-m}\left(1\pm c_{\beta}\right)^{m} $$
$$~~=(-1)^{\ell-m}2^{-\ell}\left[\displaystyle {(2\ell)! \over
(\ell+m)!(\ell-m)!}\right]^{1/2}\left(e^{i\alpha}s_{\beta}\right)^{\ell-m}
\left[e^{i\alpha}\left(c_{\gamma}+ic_{\beta}s_{\gamma}\right)
\pm e^{i\alpha}\left(c_{\beta}c_{\gamma}+is_{\gamma}\right)\right]^{m},
 \eqno (32) $$

\noindent
where $m\geq 0$. Therefore, 
$D^{\ell}_{\ell m}(\alpha, \beta, \gamma)^{*}$ can 
be expanded with respect to $Q_{q}^{\ell \tau}({\bf R}_{1},{\bf R}_{2})$,
where the coefficients depend only on the internal variables. 
$D^{\ell}_{\ell m}(\alpha,\beta,\gamma)^{*}$ constitute
a complete set of independent base functions with the angular 
momentum $\ell$, so do 
$Q_{q}^{\ell \tau}({\bf R}_{1},{\bf R}_{2})$.
Now, we come to the theorem. 

\vspace{3mm}
\noindent
{\bf Theorem}. Any function
$\Psi^{\ell \lambda}_{\ell}({\bf R}_{1}, \ldots, {\bf R}_{N-1})$
with the angular momentum ${\ell}$ and the parity
$(-1)^{\ell+\lambda}$ in a quantum $N$-body system can be
expanded with respect to the generalized harmonic polynomials 
$Q_{q}^{\ell \tau}({\bf R}_{1},{\bf R}_{2})$ with the
coefficients $\psi^{\ell \lambda}_{q \tau}(\xi,\eta,\zeta)$
depending on $(3N-6)$ internal variables: 
$$\Psi^{\ell \lambda}_{\ell}({\bf R}_{1}, \ldots, {\bf R}_{N-1})
=\displaystyle \sum_{\tau=0}^{1} \displaystyle \sum_{q=\tau}^{\ell}~
\psi^{\ell \lambda}_{q \tau}(\xi,\eta,\zeta)
Q_{q}^{\ell \tau}({\bf R}_{1},{\bf R}_{2}),  $$
$$\psi^{\ell \lambda}_{q \tau}(\xi,\eta,\zeta)=
\psi^{\ell \lambda}_{q \tau}(\xi_{1},\ldots,\xi_{N-1},\eta_{2},
\ldots, \eta_{N-1},\zeta_{3},\ldots,\zeta_{N-1}), $$
$$\psi^{\ell \lambda}_{q \tau}(\xi,\eta,-\zeta)
=(-1)^{\lambda-\tau} \psi^{\ell \lambda}_{q \tau}(\xi,\eta,\zeta),
 \eqno (33) $$

\noindent 
where the last equality means that the parity of 
$\psi^{\ell \lambda}_{q \tau}(\xi,\eta,\zeta)$ 
is $(-1)^{\lambda-\tau}$.

\section{THE GENERALIZED RADIAL EQUATIONS}

From the theorem above, the set of 
$Q_{q}^{\ell \tau}({\bf R}_{1},{\bf R}_{2})$, 
just like the set of $D^{\ell}_{\ell m}(\alpha,\beta,\gamma)^{*}$, 
is a complete set of base functions with angular momentum 
$\ell$ in the quantum $N$-body system. Each function 
with the angular momentum $\ell$ in the system
can be expanded like Eq. (33) or Eq. (2). However, Eq. (33) has 
two important characteristics, which make it easier to derive 
the generalized radial equations. One is that the generalized 
harmonic polynomial $Q_{q}^{\ell \tau}({\bf R}_{1},{\bf R}_{2})$ 
is a homogeneous polynomial in the components of two Jacobi 
coordinate vectors ${\bf R}_{1}$ and ${\bf R}_{2}$, where the 
Euler angles do not appear explicitly. The other is the well 
chosen internal variables (22), where the internal variables 
$\zeta_{j}$ have odd parity. It is due to the existence of $\zeta_{j}$ 
that $Q_{q}^{\ell 0}({\bf R}_{1},{\bf R}_{2})$ and 
$Q_{q}^{\ell 1}({\bf R}_{1},{\bf R}_{2})$ appear together in the
expansion of the wave function. By comparison, all internal 
variables in a quantum three-body system have even parity 
($\zeta_{j}=0$) so that in the expansion (20) of a wave function 
with a given parity only the base functions with the same 
parity appear \cite{ma,wig,hsi,mat}.

Because of these two characteristics, it is easy to derive 
the generalized radial equations by substituting Eq. (33) into 
the Schr\"{o}dinger equation (6) with the Laplace operator (9).
The main calculation in the derivation is to apply the Laplace
operator (9) to the function 
$\Psi^{\ell \lambda}_{\ell}({\bf R}_{1}, \ldots, {\bf R}_{N-1})$
in Eq. (33). The calculation consists of three parts. The first is to
apply the Laplace operator to the generalized radial functions
$\psi^{\ell \lambda}_{q \tau}(\xi,\eta,\zeta)$: 
$$\bigtriangleup  \psi^{\ell \lambda}_{q \tau}(\xi,\eta,\zeta)=
\left\{4\xi_{1}\partial^{2}_{\xi_{1}}+4\eta_{2}\partial^{2}_{\eta_{2}}
+\left(\xi_{1}+\eta_{2}\right)\partial^{2}_{\xi_{2}}
+4\xi_{2}\left(\partial_{\xi_{1}}
+\partial_{\eta_{2}}\right)\partial_{\xi_{2}}\right.$$
$$~~~+6\left(\partial_{\xi_{1}}+\partial_{\eta_{2}}\right)
+\displaystyle \sum_{j=3}^{N-1}~\left[\xi_{1}\partial^{2}_{\xi_{j}}
+\eta_{2}\partial^{2}_{\eta_{j}}+\Omega_{2}\partial^{2}_{\zeta_{j}}
+2\xi_{2}\partial_{\xi_{j}}\partial_{\eta_{j}}\right. $$
$$\left.~~~+4\left(\xi_{j}\partial_{\xi_{j}}
+\zeta_{j}\partial_{\zeta_{j}}\right)\partial_{\xi_{1}}
+4\left(\eta_{j}\partial_{\eta_{j}}
+\zeta_{j}\partial_{\zeta_{j}}\right)\partial_{\eta_{2}}
+2\left(\eta_{j}\partial_{\xi_{j}}
+\xi_{j}\partial_{\eta_{j}}\right)\partial_{\xi_{2}}\right]$$
$$~~~+\Omega^{-1}_{2} \displaystyle \sum_{j,k=3}^{N-1}~
\left[\left(\Omega_{j}\eta_{k}-\omega_{j}\xi_{k}
+\zeta_{j}\zeta_{k}\right) \left(\partial_{\xi_{j}}\partial_{\xi_{k}}
+\partial_{\eta_{j}}\partial_{\eta_{k}}\right) 
-2\left(\omega_{j}\zeta_{k}-\omega_{k}\zeta_{j}\right)\partial_{\xi_{j}}
\partial_{\zeta_{k}}\right.$$
$$\left.\left.~~~+2\left(\Omega_{j}\zeta_{k}-\Omega_{k}\zeta_{j}\right)
\partial_{\eta_{j}}\partial_{\zeta_{k}}
+\left(\Omega_{j}\Omega_{k}+\omega_{j}\omega_{k}
+\xi_{1}\zeta_{j}\zeta_{k}+\eta_{2}\zeta_{j}\zeta_{k}\right)
\partial_{\zeta_{j}}\partial_{\zeta_{k}} \right] 
\right\} \psi^{\ell \lambda}_{q \tau}(\xi,\eta,\zeta).  \eqno (34) $$

\noindent
The second is to apply it to the generalized harmonic polynomials 
$Q_{q}^{\ell \tau}({\bf R}_{1},{\bf R}_{2})$. This part is vanishing
because $Q_{q}^{\ell \tau}({\bf R}_{1},{\bf R}_{2})$ satisfies
the Laplace equation. The last is the mixed application
$$2\left\{\left(\partial_{\xi_{1}}\psi^{\ell \lambda}_{q \tau}\right)
2{\bf R}_{1}+\left(\partial_{\xi_{2}}\psi^{\ell \lambda}_{q \tau}\right)
{\bf R}_{2}+\displaystyle \sum_{j=3}^{N-1} \left[
\left(\partial_{\xi_{j}}\psi^{\ell \lambda}_{q \tau}\right)
{\bf R}_{j}+\left(\partial_{\zeta_{j}}\psi^{\ell \lambda}_{q \tau}\right)
\left({\bf R}_{2}\times {\bf R}_{j}\right)\right]\right\}
\cdot \bigtriangledown_{{\bf R}_{1}} Q_{q}^{\ell \tau}$$
$$+2 \left\{\left(\partial_{\xi_{2}}\psi^{\ell \lambda}_{q \tau}\right)
{\bf R}_{1}+\left(\partial_{\eta_{2}}\psi^{\ell \lambda}_{q \tau}\right)
2{\bf R}_{2}+\displaystyle \sum_{j=3}^{N-1} \left[
\left(\partial_{\eta_{j}}\psi^{\ell \lambda}_{q \tau}\right)
{\bf R}_{j}+\left(\partial_{\zeta_{j}}\psi^{\ell \lambda}_{q \tau}\right)
\left({\bf R}_{j}\times {\bf R}_{1}\right)\right]\right\}
\cdot \bigtriangledown_{{\bf R}_{2}} Q_{q}^{\ell \tau}.  $$

\noindent
In terms of Eqs. (18) and (29) we obtain: 
$${\bf R}_{1}\cdot \bigtriangledown_{{\bf R}_{1}}Q_{q}^{\ell \tau}
=qQ_{q}^{\ell \tau},~~~~~~
{\bf R}_{2}\cdot \bigtriangledown_{{\bf R}_{2}}Q_{q}^{\ell \tau}
=\left(\ell-q+\tau\right)Q_{q}^{\ell \tau}, $$
$${\bf R}_{2}\cdot \bigtriangledown_{{\bf R}_{1}}Q_{q}^{\ell \tau}
=\left(\ell-q+1\right)Q_{q-1}^{\ell \tau},~~~~~~
{\bf R}_{1}\cdot \bigtriangledown_{{\bf R}_{2}}Q_{q}^{\ell \tau}
=\left(q-\tau+1\right)Q_{q+1}^{\ell \tau},$$
$${\bf R}_{j}\cdot \bigtriangledown_{{\bf R}_{1}}Q_{q}^{\ell 0}
=\Omega_{2}^{-1}\left\{-\omega_{j}qQ_{q}^{\ell 0}
+\Omega_{j}\left(\ell-q+1\right)Q_{q-1}^{\ell 0}
-i\zeta_{j}Q_{q}^{\ell 1}\right\}, $$
$${\bf R}_{j}\cdot \bigtriangledown_{{\bf R}_{2}}Q_{q}^{\ell 0}
=\Omega_{2}^{-1}\left\{-\omega_{j}(q+1)Q_{q+1}^{\ell 0}
+\Omega_{j}\left(\ell-q\right)Q_{q}^{\ell 0}
-i\zeta_{j}Q_{q+1}^{\ell 1}\right\}, $$
$${\bf R}_{j}\cdot \bigtriangledown_{{\bf R}_{1}}Q_{q}^{\ell 1}
=\Omega_{2}^{-1}\left\{-i\eta_{2}\zeta_{j}q^{2}Q_{q}^{\ell 0}
+i\xi_{2}\zeta_{j}(2q-1)(\ell-q+1)Q_{q-1}^{\ell 0}\right.$$
$$\left.~~~-i\xi_{1}\zeta_{j}(\ell-q+2)(\ell-q+1)Q_{q-2}^{\ell 0}
-\omega_{j}qQ_{q}^{\ell 1}
+\Omega_{j}(\ell-q+1)Q_{q-1}^{\ell 1}\right\}, $$
$${\bf R}_{j}\cdot \bigtriangledown_{{\bf R}_{2}}Q_{q}^{\ell 1}
=\Omega_{2}^{-1}\left\{-i\eta_{2}\zeta_{j}(q+1)qQ_{q+1}^{\ell 0}
+i\xi_{2}\zeta_{j}q(2\ell-2q+1)Q_{q}^{\ell 0}\right.$$
$$\left.~~~-i\xi_{1}\zeta_{j}(\ell-q+1)^{2}Q_{q-1}^{\ell 0}
-\omega_{j}qQ_{q+1}^{\ell 1}
+\Omega_{j}(\ell-q+1)Q_{q}^{\ell 1}\right\},$$
$$\left({\bf R}_{2}\times {\bf R}_{j}\right)\cdot 
\bigtriangledown_{{\bf R}_{1}}Q_{q}^{\ell 0}
=\Omega_{2}^{-1}\left\{\eta_{2}\zeta_{j}qQ_{q}^{\ell 0}
-\xi_{2}\zeta_{j}\left(\ell-q+1\right)Q_{q-1}^{\ell 0}
-i\omega_{j}Q_{q}^{\ell 1}\right\}, $$
$$\left({\bf R}_{j}\times {\bf R}_{1}\right)\cdot 
\bigtriangledown_{{\bf R}_{2}}Q_{q}^{\ell 0}
=\Omega_{2}^{-1}\left\{-\xi_{2}\zeta_{j}(q+1)Q_{q+1}^{\ell 0}
+\xi_{1}\zeta_{j}\left(\ell-q\right)Q_{q}^{\ell 0}
+i\Omega_{j}Q_{q+1}^{\ell 1}\right\}, $$
$$\left({\bf R}_{2}\times {\bf R}_{j}\right)\cdot 
\bigtriangledown_{{\bf R}_{1}}Q_{q}^{\ell 1}
=\Omega_{2}^{-1}\left\{-i\eta_{2}\omega_{j}q^{2}Q_{q}^{\ell 0}
+i\xi_{2}\omega_{j}(2q-1)\left(\ell-q+1\right)Q_{q-1}^{\ell 0}\right. $$
$$\left.~~~-i\xi_{1}\omega_{j}(\ell-q+2)(\ell-q+1)Q_{q-2}^{\ell 0}
+\eta_{2}\zeta_{j}qQ_{q}^{\ell 1}
-\xi_{2}\zeta_{j}\left(\ell-q+1\right)Q_{q-1}^{\ell 1}\right\},$$
$$\left({\bf R}_{j}\times {\bf R}_{1}\right)\cdot 
\bigtriangledown_{{\bf R}_{2}}Q_{q}^{\ell 1}
=\Omega_{2}^{-1}\left\{i\eta_{2}\Omega_{j}(q+1)qQ_{q+1}^{\ell 0}
-i\xi_{2}\Omega_{j}q\left(2\ell-2q+1\right)Q_{q}^{\ell 0}\right. $$
$$\left.~~~+i\xi_{1}\Omega_{j}(\ell-q+1)^{2}Q_{q-1}^{\ell 0}
-\xi_{2}\zeta_{j}qQ_{q+1}^{\ell 1}
+\xi_{1}\zeta_{j}\left(\ell-q+1\right)Q_{q}^{\ell 1}\right\}.  \eqno (35) $$

\noindent
Now, the generalized radial equations are  
$$\bigtriangleup \psi^{\ell \lambda}_{q0} 
+4\left\{q\partial_{\xi_{1}}+(\ell-q)\partial_{\eta_{2}} \right\}
\psi^{\ell \lambda}_{q0} 
+2q\partial_{\xi_{2}} \psi^{\ell \lambda}_{(q-1)0} 
+2(\ell-q)\partial_{\xi_{2}} \psi^{\ell \lambda}_{(q+1)0} $$
$$~~~+\displaystyle \sum_{j=3}^{N-1}~2\Omega^{-1}_{2}\left\{
\left[-\omega_{j}q\partial_{\xi_{j}}
+\Omega_{j}(\ell-q)\partial_{\eta_{j}}
+\eta_{2}\zeta_{j}q\partial_{\zeta_{j}}
+\xi_{1}\zeta_{j}(\ell-q) \partial_{\zeta_{j}}\right] 
\psi^{\ell \lambda}_{q0}  \right. $$
$$~~~-q\left[\omega_{j}\partial_{\eta_{j}}
+\xi_{2}\zeta_{j}\partial_{\zeta_{j}}\right] \psi^{\ell \lambda}_{(q-1)0}
+(\ell-q)\left[\Omega_{j}\partial_{\xi_{j}}-\xi_{2}\zeta_{j}
\partial_{\zeta_{j}}\right] \psi^{\ell \lambda}_{(q+1)0} $$
$$~~~-i\eta_{2}q(q-1)\left[\zeta_{j}\partial_{\eta_{j}}
-\Omega_{j}\partial_{\zeta_{j}}\right] \psi^{\ell \lambda}_{(q-1)1}$$
$$~~~-iq\left[\eta_{2}\zeta_{j}q\partial_{\xi_{j}}
-\xi_{2}\zeta_{j}(2\ell-2q+1)\partial_{\eta_{j}}
+\eta_{2}\omega_{j}q\partial_{\zeta_{j}}
+\xi_{2}\Omega_{j}(2\ell-2q+1)\partial_{\zeta_{j}}
\right]  \psi^{\ell \lambda}_{q1} $$
$$~~~+i(\ell-q)\left[\xi_{2}\zeta_{j}(2q+1)\partial_{\xi_{j}}
-\xi_{1}\zeta_{j}(\ell-q)\partial_{\eta_{j}}
+\xi_{2}\omega_{j}(2q+1)\partial_{\zeta_{j}}
+\xi_{1}\Omega_{j}(\ell-q)\partial_{\zeta_{j}}
\right] \psi^{\ell \lambda}_{(q+1)1}$$
$$~~~\left.-i\xi_{1}(\ell-q)(\ell-q-1)\left[\zeta_{j}\partial_{\xi_{j}}
+\omega_{j}\partial_{\zeta_{j}}\right] \psi^{\ell \lambda}_{(q+2)1}\right\}
=-\left(2/\hbar^{2}\right)\left[E-V\right]\psi^{\ell \lambda}_{q0},
 \eqno (36a) $$
$$\bigtriangleup \psi^{\ell \lambda}_{q1}
+4\left\{q\partial_{\xi_{1}}+(\ell-q+1)\partial_{\eta_{2}}
\right\} \psi^{\ell \lambda}_{q1}
+2(q-1)\partial_{\xi_{2}} \psi^{\ell \lambda}_{(q-1)1}
+2(\ell-q)\partial_{\xi_{2}} \psi^{\ell \lambda}_{(q+1)1}$$
$$~~~+\displaystyle \sum_{j=3}^{N-1}~2\Omega^{-1}_{2}\left\{
\left[-\omega_{j}q\partial_{\xi_{j}}+\Omega_{j}(\ell-q+1)\partial_{\eta_{j}}
+\eta_{2}\zeta_{j}q\partial_{\zeta_{j}}+\xi_{1}\zeta_{j}(\ell-q+1)
\partial_{\zeta_{j}}\right] \psi^{\ell \lambda}_{q1}\right. $$
$$~~~-(q-1)\left[\omega_{j}\partial_{\eta_{j}}
+\xi_{2}\zeta_{j}\partial_{\zeta_{j}}
\right] \psi^{\ell \lambda}_{(q-1)1}
+(\ell-q)\left[\Omega_{j}\partial_{\xi_{j}}-\xi_{2}\zeta_{j}
\partial_{\zeta_{j}}\right] \psi^{\ell \lambda}_{(q+1)1}$$
$$~~~\left.-i\left[\zeta_{j}\partial_{\eta_{j}}
-\Omega_{j}\partial_{\zeta_{j}}\right] \psi^{\ell \lambda}_{(q-1)0}
-i\left[\zeta_{j}\partial_{\xi_{j}}
+\omega_{j}\partial_{\zeta_{j}}\right] \psi^{\ell \lambda}_{q0}\right\}
=-\left(2/\hbar^{2}\right)\left[E-V\right] \psi^{\ell \lambda}_{q1},
 \eqno (36b) $$

\noindent
where $\bigtriangleup \psi^{\ell \lambda}_{q\tau}$ was given in 
Eq. (34). When $N$=3, Eq. (36) reduces to Eq. (21), 
where, because all internal variables have even parity, the 
generalized radial functions 
$\psi^{\ell \lambda}_{q \tau}(\xi,\eta,\zeta)$ with
$\lambda\neq \tau$ have to be vanishing. 

\section{PERMUTATION PROPERTY OF WAVE FUNCTIONS}

When some or all particles in a quantum $N$-body system
are identical particles, one has to consider the
permutation property of the spatial wave function, 
which depends on the total spin of identical particles. 
Since the spatial wave function 
$\Psi^{\ell \lambda}_{\ell}({\bf R}_{1}, \ldots, {\bf R}_{N-1})$
is expanded with respect to the base functions
$Q_{q}^{\ell \tau}({\bf R}_{1},{\bf R}_{2})$, we need to study 
only the property of $Q_{q}^{\ell \tau}({\bf R}_{1},{\bf R}_{2})$
in the transposition $(k,k+1)$ between two neighboring particles.
The transformation property of the Jacobi coordinate vectors
${\bf R}_{j}$ in the transposition $(k,k+1)$ was given in
Eq. (10). Therefore, the base function
$Q_{q}^{\ell \tau}({\bf R}_{1},{\bf R}_{2})$ remains invariant
in the transposition $(k,k+1)$ with $k\geq 3$. In the following
we are going to study the transformation property of
$Q_{q}^{\ell \tau}({\bf R}_{1},{\bf R}_{2})$ in the transpositions
$(1,2)$ and $(2,3)$. Denote by $P_{1}$ and $P_{2}$ the transformation
operators for the base function in the transpositions $(1,2)$ 
and $(2,3)$, respectively. In the following formulas we
neglect the argument ${\bf R}_{1},{\bf R}_{2}$ in 
$Q_{q}^{\ell \tau}({\bf R}_{1},{\bf R}_{2})$ and briefly denote
$\sin \theta_{1}$, $\sin \theta_{2}$, $\cos \theta_{1}$, and 
$\cos \theta_{2}$ by $s_{1}$, $s_{2}$, $c_{1}$, and $c_{2}$
for simplicity. 

\begin{center}
{\bf A. Transposition} $(1,2)$
\end{center}

$$P_{1}Q_{q}^{\ell 0}=\displaystyle {1 \over q!(\ell-q)!}
\left[-Xc_{1}+Ys_{1}\right]^{q}\left[Xs_{1}+Yc_{1}\right]^{\ell-q}
=\displaystyle \sum_{p=0}^{\ell}~
Q_{p}^{\ell 0}D^{\ell (1)}_{pq}(\theta_{1}). 
\eqno (37) $$

\noindent
where
$$D^{\ell (1)}_{pq}(\theta_{1})=\displaystyle \sum_{n}~
\displaystyle {(-1)^{q-n}p!(\ell-p)! c_{1}^{\ell-p+q-2n} s_{1}^{2n+p-q}
\over (q-n)!(\ell-p-n)!n!(n+p-q)!}.  \eqno (38) $$

\noindent
Because of Eq. (19) and $P_{1}Z=-Z$, we obtain 
$$P_{1}Q_{q}^{\ell 1}= - \displaystyle \sum_{p=1}^{\ell}~
Q_{p}^{\ell 1}D^{(\ell-1) (1)}_{(p-1)(q-1)}(\theta_{1}). \eqno (39) $$

\begin{center}
{\bf B. Transposition} $(2,~3)$
\end{center}

$$P_{2}Q_{q}^{\ell 0}=\displaystyle {X^{q} 
\left[-s_{2}\omega_{3}X+\left(s_{2}\Omega_{3}-c_{2}\Omega_{2}\right)Y
-is_{2}\zeta_{3}Z\right]^{\ell-q}
\over q!(\ell-q)! \Omega_{2}^{\ell-q}}$$
$$=\displaystyle \sum_{p=q}^{\ell}~
Q_{p}^{\ell 0}D^{\ell (2)}_{pq}(\theta_{2})
+i\displaystyle \sum_{p=q+1}^{\ell}~
Q_{p}^{\ell 1}D^{\ell (3)}_{pq}(\theta_{2}), \eqno (40) $$

\noindent
where
$$D^{\ell (2)}_{(q+n)q}=\displaystyle {(q+n)!(\ell-q-n)!s_{2}^{\ell-q}
\over q!(\ell-q)! \Omega_{2}^{\ell-q}} 
\displaystyle \sum_{m}~\left(\begin{array}{c}
\ell-q \\ 2m \end{array} \right) (-1)^{n+m}\zeta_{3}^{2m}$$
$$~~~\times \displaystyle \sum_{rt}~
\left(\begin{array}{c} \ell-q-2m \\ n-r \end{array} \right)
\left(\begin{array}{c} m \\ t \end{array} \right)
\left(\begin{array}{c} m-t \\ r-2t \end{array} \right)$$
$$~~~\times \xi_{1}^{m-r+t}\eta_{2}^{t}\left(2\xi_{2}\right)^{r-2t}
\omega_{3}^{n-r}\left(\Omega_{3}
-\Omega_{2}c_{2}/s_{2}\right)^{\ell-q-n-2m+r}, $$
$$D^{\ell (3)}_{(q+n)q}=\displaystyle {(q+n-1)!(\ell-q-n)!s_{2}^{\ell-q}
\over q!(\ell-q)! \Omega_{2}^{\ell-q}}
\displaystyle \sum_{m}~\left(\begin{array}{c}
\ell-q \\ 2m+1 \end{array} \right) (-1)^{n+m}\zeta_{3}^{2m+1} $$
$$~~~\times \displaystyle \sum_{rt}~
\left(\begin{array}{c} \ell-q-2m-1 \\ n-r-1 \end{array} \right)
\left(\begin{array}{c} m \\ t \end{array} \right)
\left(\begin{array}{c} m-t \\ r-2t \end{array} \right) $$
$$~~~\times \xi_{1}^{m-r+t}\eta_{2}^{t}\left(2\xi_{2}\right)^{r-2t}
\omega_{3}^{n-r-1}\left(\Omega_{3}
-\Omega_{2}c_{2}/s_{2}\right)^{\ell-q-n-2m+r},  \eqno (41) $$

\noindent
where the combinatorics 
$\left(\begin{array}{c} a \\[-2mm] b \end{array} \right)
=\displaystyle {a! \over b!(a-b)!}$, and the ranges of the 
summation indices $m$, $r$ and $t$ are determined by the
conditions that those combinatorics are not vanishing. 
$$P_{2}Q_{q}^{\ell 1}=\Omega_{2}^{-1}
\left\{is_{2}\xi_{2}\zeta_{3}X-is_{2}\xi_{1}\zeta_{3}Y
+\left(s_{2}\Omega_{3}-c_{2}\Omega_{2}\right)Z\right\}
\left[P_{2}Q_{q-1}^{(\ell-1)0}\right]$$
$$=is_{2}\zeta_{3}\Omega_{2}^{-1}\displaystyle \sum_{p=q}^{\ell}~
\left\{Q_{p}^{\ell 0}p\xi_{2}-Q_{p-1}^{\ell 0}(\ell-p+1)\xi_{1}\right\}
D^{(\ell-1) (2)}_{(p-1)(q-1)}(\theta_{2})$$
$$~~~+i\left(s_{2}\Omega_{3}-c_{2}\Omega_{2}\right)\Omega_{2}^{-1}
\displaystyle \sum_{p=q+1}^{\ell}~\left\{Q_{p}^{\ell 0}
p(p-1)\eta_{2}-2Q_{p-1}^{\ell 0}(p-1)(\ell-p+1)\xi_{2}\right. $$
$$~~~\left. +Q_{p-2}^{\ell 0}(\ell-p+2)(\ell-p+1)\xi_{1}\right\}
D^{(\ell-1) (3)}_{(p-1)(q-1)}(\theta_{2})$$
$$~~~+\Omega_{2}^{-1}\left(s_{2}\Omega_{3}-c_{2}\Omega_{2}\right)
\displaystyle \sum_{p=q}^{\ell}~Q_{p}^{\ell 1}
D^{(\ell-1) (2)}_{(p-1)(q-1)}(\theta_{2})$$
$$~~~-s_{2}\zeta_{3}\Omega_{2}^{-1}\displaystyle \sum_{p=q+1}^{\ell}~
\left\{Q_{p}^{\ell 1}(p-1)\xi_{2}-Q_{p-1}^{\ell 1}(\ell-p+1)\xi_{1}\right\}
D^{(\ell-1) (3)}_{(p-1)(q-1)}(\theta_{2}).  \eqno (42) $$

In real calculations the cases with the small angular momentum
may be more interesting. In the following we explicitly list
the above formulas for $\ell=1$ and $2$ (the formulas for the case 
with $\ell=0$ are trivial):
$$P_{1}Q_{1}^{10}=-c_{1}Q_{1}^{10}+s_{1}Q_{0}^{10},~~~~~
P_{1}Q_{0}^{10}=s_{1}Q_{1}^{10}+c_{1}Q_{0}^{10},~~~~~
P_{1}Q_{1}^{11}=-Q_{1}^{11}, $$
$$P_{1}Q_{2}^{20}=c_{1}^{2}Q_{2}^{20}-c_{1}s_{1}Q_{1}^{20}
+s_{1}^{2}Q_{0}^{20},$$
$$P_{1}Q_{1}^{20}=-2c_{1}s_{1}Q_{2}^{20}
-\left(c_{1}^{2}-s_{1}^{2}\right)Q_{1}^{20}
+2c_{1}s_{1}Q_{0}^{20},$$
$$P_{1}Q_{0}^{20}=s_{1}^{2}Q_{2}^{20}+c_{1}s_{1}Q_{1}^{20}
+c_{1}^{2}Q_{0}^{20},$$
$$P_{1}Q_{2}^{21}=c_{1}Q_{2}^{21}-s_{1}Q_{1}^{21},~~~~~
P_{1}Q_{1}^{21}=-s_{1}Q_{2}^{21}-c_{1}Q_{1}^{21}.  $$
$$P_{2}Q_{1}^{10}=Q_{1}^{10},~~~~~
P_{2}Q_{0}^{10}=s_{2}\Omega_{2}^{-1}\left[-\omega_{3}Q_{1}^{10}
+\left(\Omega_{3}-\Omega_{2}c_{2}/s_{2}\right)Q_{0}^{10}
-i\zeta_{3}Q_{1}^{11} \right], $$
$$P_{2}Q_{1}^{11}=s_{2}\Omega_{2}^{-1}\left[i\xi_{2}\zeta_{3}Q_{1}^{10}
-i\xi_{1}\zeta_{3}Q_{0}^{10}
+\left(\Omega_{3}-\Omega_{2}c_{2}/s_{2}\right)Q_{1}^{11}\right], $$
$$P_{2}Q_{2}^{20}=Q_{2}^{20},~~~~~
P_{2}Q_{1}^{20}=s_{2}\Omega_{2}^{-1}\left[-2\omega_{3}Q_{2}^{20}
+\left(\Omega_{3}-\Omega_{2}c_{2}/s_{2}\right)Q_{1}^{20}
-i\zeta_{3}Q_{2}^{21} \right], $$
$$P_{2}Q_{0}^{20}=s_{2}^{2}\Omega_{2}^{-2}\left\{\left(\omega_{3}^{2}
-\eta_{2}\zeta_{3}^{2}\right)Q_{2}^{20}
+\left[-\omega_{3}\left(\Omega_{3}-\Omega_{2}c_{2}/s_{2}\right)
+\xi_{2}\zeta_{3}^{2}\right]Q_{1}^{20} \right.$$
$$~~~\left.+\left[\left(\Omega_{3}-\Omega_{2}c_{2}/s_{2}\right)^{2}
-\xi_{1}\zeta_{3}^{2}\right]Q_{0}^{20}
+i\omega_{3}\zeta_{3}Q_{2}^{21}
-i\zeta_{3}\left(\Omega_{3}-\Omega_{2}c_{2}/s_{2}\right)Q_{1}^{21}\right\},$$
$$P_{2}Q_{2}^{21}=s_{2}\Omega_{2}^{-1}\left\{i2\xi_{2}\zeta_{3}Q_{2}^{20}
-i\xi_{1}\zeta_{3}Q_{1}^{20}
+\left(\Omega_{3}-\Omega_{2}c_{2}/s_{2}\right)Q_{2}^{21}\right\}, $$
$$P_{2}Q_{1}^{21}=s_{2}^{2}\Omega_{2}^{-2}\left\{-2i\zeta_{3}\left[
\xi_{2}\omega_{3}+\eta_{2}\left(\Omega_{3}-\Omega_{2}c_{2}/s_{2}\right)
\right]Q_{2}^{20}\right. $$
$$~~~+i\zeta_{3}\left[\xi_{1}\omega_{3}
+3\xi_{2}\left(\Omega_{3}-\Omega_{2}c_{2}/s_{2}\right)\right]Q_{1}^{20}
-4i\xi_{1}\zeta_{3}\left(\Omega_{3}-\Omega_{2}c_{2}/s_{2}\right)Q_{0}^{20}$$
$$~~~\left.+\left[\xi_{2}\zeta_{3}^{2}-\omega_{3}
\left(\Omega_{3}-\Omega_{2}c_{2}/s_{2}\right)\right]Q_{2}^{21}
+\left[\left(\Omega_{3}-\Omega_{2}c_{2}/s_{2}\right)^{2}
-\xi_{1}\zeta_{3}^{2}\right]Q_{1}^{21}\right\}.  $$

\section{NONORTHOGONAL VECTORS}

Now, we turn to the general case where arbitrary coordinate vectors
${\bf r}_{cj}$ in the center-of-mass frame [see Eq. (13)] are used 
to replace the Jacobi coordinate vectors ${\bf R}_{j}$. In this case 
the Laplace operator contains mixed derivative terms [see Eq. (14)]. 
All the conclusions in Sec. III hold for the present case except 
that the Jacobi coordinate vectors should be replaced with the 
coordinate vectors ${\bf r}_{cj}$ and the volume element of the 
configuration space (25) changes due to the linear transformation 
(13). In particular, the generalized harmonic polynomial 
$Q^{\ell \lambda}_{q}({\bf R}_{1},{\bf R}_{2})$ 
now becomes $Q^{\ell \lambda}_{q}({\bf r}_{c1},{\bf r}_{c2})$, where
${\bf r}_{c1}$ and ${\bf r}_{c2}$ are two arbitrarily chosen  
coordinate vectors. 

Any function 
$\Psi^{\ell \lambda}_{\ell}({\bf r}_{c1},\ldots,{\bf r}_{c(N-1)})$
with angular momentum $\ell$ and parity $(-1)^{\ell+\lambda}$
in a quantum $N$-body system can be expanded with respect to 
$Q^{\ell \lambda}_{q}({\bf r}_{c1},{\bf r}_{c2})$ with the coefficients
$\psi^{\ell \lambda}_{q \tau}(\xi,\eta,\zeta)$ depending on $(3N-6)$
invariant variables
$$\Psi^{\ell \lambda}_{\ell}({\bf r}_{c1}, \ldots, {\bf r}_{c(N-1)})
=\displaystyle \sum_{\tau=0}^{1} \displaystyle \sum_{q=\tau}^{\ell}~
\psi^{\ell \lambda}_{q \tau}(\xi,\eta,\zeta)
Q_{q}^{\ell \tau}({\bf r}_{c1},{\bf r}_{c2}),  \eqno (43) $$

\noindent
where, instead of Eq. (22), the internal variables $\xi_{j}$, 
$\eta_{j}$, and $\zeta_{j}$ are redefined as 
$$\xi_{j}={\bf r}_{cj}\cdot {\bf r}_{c1},~~~~~
\eta_{j}={\bf r}_{cj}\cdot {\bf r}_{c2},~~~~~
\zeta_{j}={\bf r}_{cj}\cdot \left({\bf r}_{c1} \times 
{\bf r}_{c2} \right),$$
$$1\leq j \leq (N-1),~~~~~~\eta_{1}=\xi_{2},~~~~~~
\zeta_{1}=\zeta_{2}=0.  \eqno (44) $$

As in the case with the orthogonal vectors, the main calculation 
in deriving the generalized radial equations in the present case
is to apply the Laplace operator (14) to the function 
$\Psi^{\ell \lambda}_{\ell}({\bf r}_{c1}, \ldots, {\bf r}_{c(N-1)})$
in Eq. (43). Similarly, the calculation consists of three parts,
and the second part is vanishing. But, the first part [see Eq. (34)]
becomes 
$$\bigtriangleup  \psi^{\ell \lambda}_{q \tau}(\xi,\eta,\zeta)
=\left\{(4S_{11}\xi_{1}\partial_{\xi_{1}}^{2}
+4S_{22}\eta_{2}\partial_{\eta_{2}}^{2})
+\left(S_{11}\eta_{2}+S_{22}\xi_{1}+2 S_{12}\xi_{2}\right)
\partial_{\xi_{2}}^{2}  \right.$$
$$+4\left(S_{11}\xi_{2}+S_{12}\xi_{1}\right)
\partial_{\xi_{1}}\partial_{\xi_{2}}
+4\left(S_{22}\xi_{2}+S_{12}\eta_{2}\right)
\partial_{\xi_{2}}\partial_{\eta_{2}}$$
$$\left.~~~+8 S_{12}\xi_{2}\partial_{\xi_{1}}\partial_{\eta_{2}}
+6\left(S_{11}\partial_{\xi_{1}}+ S_{22}\partial_{\eta_{2}}
+ S_{12}\partial_{\xi_{2}}\right)\right\}
\psi^{\ell \lambda}_{q \tau}(\xi,\eta,\zeta) $$
$$~~~+\displaystyle \sum_{j=3}^{N-1}\left\{
4 \left(S_{11}\xi_{j}+S_{1j}\xi_{1}\right)
\partial_{\xi_{1}}\partial_{\xi_{j}}
+4 \left(S_{22}\eta_{j}+S_{2j}\eta_{2}\right)
\partial_{\eta_{2}}\partial_{\eta_{j}}
+4 \left(S_{12}\xi_{j}+S_{1j}\xi_{2}\right)
\partial_{\xi_{1}}\partial_{\eta_{j}}\right. $$
$$~~~+4 \left(S_{12}\eta_{j}+S_{2j}\xi_{2}\right)
\partial_{\eta_{2}}\partial_{\xi_{j}}
+2 \left(S_{11}\eta_{j}+S_{12}\xi_{j}+S_{1j}\xi_{2}+S_{2j}\xi_{1}\right)
\partial_{\xi_{2}}\partial_{\xi_{j}}$$
$$~~~+2 \left(S_{22}\xi_{j}+S_{12}\eta_{j}+S_{1j}\eta_{2}+S_{2j}\xi_{2}\right)
\partial_{\xi_{2}}\partial_{\eta_{j}}
+4 \zeta_{j} \left(S_{11}\partial_{\xi_{1}}+S_{22}\partial_{\eta_{2}}
+S_{12} \partial_{\xi_{2}}\right)\partial_{\zeta_{j}}$$
$$\left.~~~+6\left(S_{1j}\partial_{\xi_{j}}+S_{2j}\partial_{\eta_{j}}\right)
\right\} \psi^{\ell \lambda}_{q \tau}(\xi,\eta,\zeta)
+\displaystyle \sum_{j,k=3}^{N-1}~\left\{
\left(2S_{1j}\xi_{k}+S_{jk}\xi_{1}\right)
\partial_{\xi_{j}}\partial_{\xi_{k}} \right. $$
$$~~~+\left(2S_{2j}\eta_{k}+S_{jk}\eta_{2}\right)
\partial_{\eta_{j}}\partial_{\eta_{k}}
+2\left(S_{1k}\eta_{j}+S_{2j}\xi_{k}+S_{jk}\xi_{2}\right)
\partial_{\xi_{j}}\partial_{\eta_{k}}
+2\left(S_{1j}\zeta_{k}+S_{1k}\zeta_{j}\right)
\partial_{\xi_{j}}\partial_{\zeta_{k}} $$
$$\left.~~~+2\left(S_{2j}\zeta_{k}+S_{2k}\zeta_{j}\right)
\partial_{\eta_{j}}\partial_{\zeta_{k}}
+\left(2S_{1j}\omega_{k}-2S_{2j}\Omega_{k}+S_{jk}\Omega_{2}\right)
\partial_{\zeta_{j}}\partial_{\zeta_{k}}
\right\} \psi^{\ell \lambda}_{q \tau}(\xi,\eta,\zeta)$$
$$~~~+\displaystyle \sum_{j,k=3}^{N-1}~\Omega_{2}^{-1}\left\{
\left(\Omega_{j}\eta_{k}-\omega_{j}\xi_{k}+\zeta_{j}\zeta_{k}\right)
\left(S_{11}\partial_{\xi_{j}}\partial_{\xi_{k}}
+S_{22}\partial_{\eta_{j}}\partial_{\eta_{k}}
+2S_{12}\partial_{\xi_{j}}\partial_{\eta_{k}}\right)\right.$$
$$~~~+2\left[-S_{11}\left(\omega_{j}\zeta_{k}-\omega_{k}\zeta_{j}\right)
+S_{12}\left(\Omega_{j}\zeta_{k}-\Omega_{k}\zeta_{j}\right)\right]
\partial_{\xi_{j}}\partial_{\zeta_{k}}$$
$$~~~+2\left[S_{22}\left(\Omega_{j}\zeta_{k}-\Omega_{k}\zeta_{j}\right)
-S_{12}\left(\omega_{j}\zeta_{k}-\omega_{k}\zeta_{j}\right)\right]
\partial_{\eta_{j}}\partial_{\zeta_{k}}
+\left[S_{11}\left(\omega_{j}\omega_{k}
+\eta_{2}\zeta_{j}\zeta_{k}\right)\right.$$
$$\left.\left.~~~+S_{22}\left(\Omega_{j}\Omega_{k}+\xi_{1}\zeta_{j}\zeta_{k}\right)
-2S_{12}\left(\omega_{j}\Omega_{k}+\xi_2 \zeta_{j}\zeta_{k}\right)\right]
\partial_{\zeta_{j}}\partial_{\zeta_{k}}
\right\} \psi^{\ell \lambda}_{q \tau}(\xi,\eta,\zeta). \eqno (45) $$

\noindent
The last part contains the mixed application
$$2\displaystyle \sum_{j=1}^{N-1}\displaystyle \sum_{k=1}^{N-1}~
S_{jk}\bigtriangledown_{{\bf r}_{cj}}
\psi^{\ell \lambda}_{q\tau}\cdot \bigtriangledown_{{\bf r}_{ck}}
Q^{\ell \tau}_{q}~~~~~~~~~~~~~~~~~~~~~~~~~~~~~~~~~~~~~~~~~~~~~~~~~~~~$$
$$=2\left\{{\bf r}_{c1}\left[2S_{11}\partial_{\xi_{1}}+S_{12}\partial_{\xi_{2}}
+\displaystyle \sum_{j=3}^{N-1}~S_{1j}\partial_{\xi_{j}}\right]
+{\bf r}_{c2}\left[S_{11}\partial_{\xi_{2}}+2S_{12}\partial_{\eta_{2}}
+\displaystyle \sum_{j=3}^{N-1}~S_{1j}\partial_{\eta_{j}}\right]\right.$$
$$~~~ +\displaystyle \sum_{j=3}^{N-1}~\left[{\bf r}_{cj}
\left(S_{11}\partial_{\xi_{j}}+S_{12}\partial_{\eta_{j}}\right)
+\left({\bf r}_{c2}\times {\bf r}_{cj}\right) S_{11}\partial_{\zeta_{j}}
+\left({\bf r}_{cj}\times {\bf r}_{c1}\right) S_{12}\partial_{\zeta_{j}}
\right. $$
$$\left.\left.~~~+\left({\bf r}_{c1}\times {\bf r}_{c2}\right) S_{1j}
\partial_{\zeta_{j}}\right]\right\}
\psi^{\ell \lambda}_{q \tau} \cdot \bigtriangledown_{{\bf r}_{c1}}
Q^{\ell \tau}_{q} 
+2\left\{{\bf r}_{c1}\left[2S_{12}\partial_{\xi_{1}}
+S_{22}\partial_{\xi_{2}}
+\displaystyle \sum_{j=3}^{N-1}~S_{2j}\partial_{\xi_{j}}\right]\right.$$
$$~~~+{\bf r}_{c2}\left[S_{12}\partial_{\xi_{2}}+2S_{22}\partial_{\eta_{2}}
+\displaystyle \sum_{j=3}^{N-1}~S_{2j}\partial_{\eta_{j}}\right] 
 +\displaystyle \sum_{j=3}^{N-1}~\left[{\bf r}_{cj}
\left(S_{12}\partial_{\xi_{j}}+S_{22}\partial_{\eta_{j}}\right)
\right. $$
$$\left.\left.~~~+\left({\bf r}_{c2}\times {\bf r}_{cj}\right) S_{12}
\partial_{\zeta_{j}}
+\left({\bf r}_{cj}\times {\bf r}_{c1}\right) S_{22}\partial_{\zeta_{j}}
+\left({\bf r}_{c1}\times {\bf r}_{c2}\right) 
S_{2j}\partial_{\zeta_{j}} \right]\right\}
\psi^{\ell \lambda}_{q \tau} \cdot \bigtriangledown_{{\bf r}_{c2}}
Q^{\ell \tau}_{q} . \eqno (46) $$

\noindent
In addition to the formulas (35), where ${\bf R}_{j}$ should be 
replaced with ${\bf r}_{cj}$, we also need the following formulas
$$\left({\bf r}_{c2}\times {\bf r}_{cj}\right)\cdot
\bigtriangledown_{{\bf r}_{c2}}Q_{q}^{\ell 0}
=\Omega_{2}^{-1}\left\{\eta_{2}\zeta_{j}(q+1)Q_{q+1}^{\ell 0}
-\xi_{2}\zeta_{j}(\ell-q)Q_{q}^{\ell 0}
-i\omega_{j}Q_{q+1}^{\ell 1}\right\},$$
$$\left({\bf r}_{cj}\times {\bf r}_{c1}\right)\cdot
\bigtriangledown_{{\bf r}_{c1}}Q_{q}^{\ell 0}
=\Omega_{2}^{-1}\left\{-\xi_{2}\zeta_{j}q Q_{q}^{\ell 0}
+\xi_{1}\zeta_{j}(\ell-q+1)Q_{q-1}^{\ell 0}
+i\Omega_{j}Q_{q}^{\ell 1}\right\},$$
$$\left({\bf r}_{c2}\times {\bf r}_{cj}\right)\cdot
\bigtriangledown_{{\bf r}_{c2}}Q_{q}^{\ell 1}
=\Omega_{2}^{-1}\left\{-i\xi_{1}\omega_{j}(\ell-q+1)^{2}Q_{q-1}^{\ell 0}
+i\xi_{2}\omega_{j} q \left(2\ell-2q+1\right)Q_{q}^{\ell 0}\right. $$
$$\left.~~~-i\eta_{2}\omega_{j}(q+1)q Q_{q+1}^{\ell 0}
-\xi_{2}\zeta_{j}(\ell-q+1) Q_{q}^{\ell 1}
+\eta_{2}\zeta_{j} q Q_{q+1}^{\ell 1}\right\},$$
$$\left({\bf r}_{cj}\times {\bf r}_{c1}\right)\cdot
\bigtriangledown_{{\bf r}_{c1}}Q_{q}^{\ell 1}
=\Omega_{2}^{-1}\left\{i\xi_{1}\Omega_{j}(\ell-q+1)(\ell-q+2)Q_{q-2}^{\ell 0}
\right. $$
$$\left.~~~-i\xi_{2}\Omega_{j}  
\left(\ell-q+1\right)(2q-1)Q_{q-1}^{\ell 0}
+i\eta_{2}\Omega_{j}q^{2} Q_{q}^{\ell 0}
+\xi_{1}\zeta_{j}(\ell-q+1) Q_{q-1}^{\ell 1}
-\xi_{2}\zeta_{j} q Q_{q}^{\ell 1}\right\},$$
$$\left({\bf r}_{c1}\times {\bf r}_{c2}\right)\cdot
\bigtriangledown_{{\bf r}_{c1}}Q_{q}^{\ell 0}
=-i Q_{q}^{\ell 1},$$
$$\left({\bf r}_{c1}\times {\bf r}_{c2}\right)\cdot
\bigtriangledown_{{\bf r}_{c2}}Q_{q}^{\ell 0}
=-i Q_{q+1}^{\ell 1},$$
$$\left({\bf r}_{c1}\times {\bf r}_{c2}\right)\cdot
\bigtriangledown_{{\bf r}_{c1}}Q_{q}^{\ell 1}
=-i \eta_{2} q^{2} Q_{q}^{\ell 0}
+i \xi_{2}(\ell-q+1)(2q-1)Q_{q-1}^{\ell 0}
-i\xi_{1}(\ell-q+2)(\ell-q+1) Q_{q-2}^{\ell 0},$$
$$\left({\bf r}_{c1}\times {\bf r}_{c2}\right)\cdot
\bigtriangledown_{{\bf r}_{c2}}Q_{q}^{\ell 1}
=i \xi_{2} q (2\ell-2q+1) Q_{q}^{\ell 0}
-i \xi_{1}(\ell-q+1)^{2} Q_{q-1}^{\ell 0}
-i\eta_{2} q(q+1) Q_{q+1}^{\ell 0}, \eqno (47) $$

\noindent
Finally, we obtain the generalized radial equations as follows
$$\bigtriangleup \psi^{\ell \lambda}_{q0}
+2\left[2 S_{11}q\partial_{\xi_{1}}+\ell S_{12}\partial_{\xi_{2}}
+2(\ell-q) S_{22}\partial_{\eta_{2}}
\right] \psi^{\ell  \lambda}_{q0}$$
$$+2 q \left(2 S_{12} \partial_{\xi_{1}}+S_{22}\partial_{\xi_{2}}\right)
\psi^{\ell \lambda}_{(q-1)0}
+2(\ell-q)\left(S_{11}\partial_{\xi_{2}}+ 2 S_{12} \partial_{\eta_{2}}\right)
\psi^{\ell \lambda}_{(q+1)0}$$
$$+\displaystyle \sum_{j=3}^{N-1}~2\Omega^{-1}_{2}\left\{
\left[ (\ell-q) \left(\Omega_{j}S_{12}\partial_{\xi_{j}}
+\Omega_{j} S_{22}\partial_{\eta_{j}}
+\Omega_{2} S_{j2}\partial_{\eta_{j}}
+\xi_{1} \zeta_{j} S_{22}\partial_{\zeta_{j}}\right)\right.\right.$$
$$\left.+q \left(\Omega_{2} S_{j1}-\omega_{j} S_{11}\right)\partial_{\xi_{j}}
-\omega_{j} q S_{12}\partial_{\eta_{j}}
+\zeta_{j}\left(\eta_{2} q S_{11}-\xi_{2} \ell S_{12}
\right)\partial_{\zeta_{j}}\right] 
\psi^{\ell \lambda}_{q0} $$
$$+q\left[\left(\Omega_{2} S_{j2}-\omega_{j} S_{12}\right)
\partial_{\xi_{j}}
-\omega_{j} S_{22}\partial_{\eta_{j}}
+\zeta_{j} \left(\eta_{2}S_{12}-\xi_{2}S_{22}\right)
\partial_{\zeta_{j}}\right]
\psi^{\ell \lambda}_{(q-1)0}$$
$$+(\ell-q)\left[S_{11}\Omega_{j}\partial_{\xi_{j}}
+\left(S_{j1}\Omega_{2} +S_{12}\Omega_{j}\right)\partial_{\eta_{j}}
+\zeta_{j} \left(\xi_{1} S_{12}-\xi_{2}S_{11}\right)
\partial_{\zeta_{j}}\right]\psi^{\ell \lambda}_{(q+1)0}$$
$$-i\eta_{2} q (q-1)\left\{
\zeta_{j} \left(S_{12}\partial_{\xi_{j}}+S_{22}\partial_{\eta_{j}}\right)
+\left(\omega_{j} S_{12}-\Omega_{j} S_{22}+\Omega_{2} S_{j2}\right)
\partial_{\zeta_{j}}\right\}\psi^{\ell \lambda}_{(q-1) 1}$$
$$-i q \left[\eta_{2} q \left(
 \zeta_{j} S_{11}\partial_{\xi_{j}}
+ \zeta_{j} S_{12}\partial_{\eta_{j}}
+ \omega_{j} S_{11}\partial_{\zeta_{j}}
-\Omega_{j} S_{12} \partial_{\zeta_{j}}
+ \Omega_{2} S_{j1}\partial_{\zeta_{j}}\right)\right.$$
$$\left.-\xi_{2} (2\ell-2q+1) \left(\zeta_{j}  S_{12}\partial_{\xi_{j}}
+ \zeta_{j} S_{22}\partial_{\eta_{j}}
+\Omega_{2}  S_{j2}\partial_{\zeta_{j}}
-\Omega_{j} S_{22}\partial_{\zeta_{j}}
+\omega_{j} S_{12}\partial_{\zeta_{j}}\right)
\right]\psi^{\ell \lambda}_{q 1}$$
$$+i(\ell-q)\left[(2q+1)\xi_{2}\left( \zeta_{j} S_{11}\partial_{\xi_{j}}
+ \zeta_{j} S_{12}\partial_{\eta_{j}}
+\Omega_{2}  S_{j1}\partial_{\zeta_{j}}
+\omega_{j} S_{11}\partial_{\zeta_{j}}
-\Omega_{j} S_{12}\partial_{\zeta_{j}}\right)\right. $$
$$\left.-\xi_{1} (\ell-q) \left(\zeta_{j}S_{12}\partial_{\xi_{j}}
+\zeta_{j}S_{22}\partial_{\eta_{j}}
+ \omega_{j} S_{12}\partial_{\zeta_{j}}
- \Omega_{j} S_{22}\partial_{\zeta_{j}}
+ \Omega_{2} S_{j2}\partial_{\zeta_{j}}\right)
\right]\psi^{\ell \lambda}_{(q+1) 1}$$
$$\left.-i \xi_{1}(\ell-q)(\ell-q-1)\left[ \zeta_{j}\left(S_{11}
\partial_{\xi_{j}}+S_{12}\partial_{\eta_{j}}\right)
+\left(\omega_{j} S_{11}-\Omega_{j} S_{12}+\Omega_{2} S_{j1}\right)
\partial_{\zeta_{j}}\right]\psi^{\ell \lambda}_{(q+2) 1}\right\}$$
$$=-\left(2/\hbar^{2}\right)\left[E-V\right] \psi^{\ell \lambda}_{q0},
 \eqno (48a) $$
$$\bigtriangleup \psi^{\ell \lambda}_{q1}
+2\left\{2 q S_{11}\partial_{\xi_{1}}+(\ell+1) S_{12}\partial_{\xi_{2}}
+2(\ell-q+1) S_{22}\partial_{\eta_{2}}
\right\} \psi^{\ell  \lambda}_{q1}$$
$$+2(q-1)\left(2 S_{12} \partial_{\xi_{1}}+S_{22}\partial_{\xi_{2}}\right)
\psi^{\ell \lambda}_{(q-1)1}
+2(\ell-q)\left(S_{11}\partial_{\xi_{2}}+ 2 S_{12} \partial_{\eta_{2}}\right)
\psi^{\ell \lambda}_{(q+1)1}$$
$$+\displaystyle \sum_{j=3}^{N-1}~2\Omega^{-1}_{2}\left\{
-i\left[\zeta_{j}\left(S_{12}\partial_{\xi_{j}}+S_{22}\partial_{\eta_{j}}
\right)+\left(\Omega_{2}S_{j2}-\Omega_{j}S_{22}+\omega_{j}S_{12}\right)
\partial_{\zeta_{j}}\right]
\psi^{\ell \lambda}_{(q-1)0}\right. $$
$$-i\left[\zeta_{j}\left(S_{11}\partial_{\xi_{j}}+S_{12}\partial_{\eta_{j}}
\right)+\left(\omega_{j} S_{11}-\Omega_{j} S_{12}+ \Omega_{2}S_{j1}\right)
\partial_{\zeta_{j}}\right] \psi^{\ell \lambda}_{q0}$$
$$+\left[q \left(\Omega_{2} S_{j1}-\omega_{j} S_{11}\right)
\partial_{\xi_{j}}
-\omega_{j} q S_{12} \partial_{\eta_{j}}
+\zeta_{j}\left(\eta_{2} q S_{11}-\xi_{2} (\ell+1)S_{12}\right)
\partial_{\zeta_{j}} \right. $$
$$\left.+(\ell-q+1)\left(\Omega_{j}S_{12} \partial_{\xi_{j}}
+\Omega_{2}S_{j2}\partial_{\eta_{j}} +\Omega_{j}S_{22}\partial_{\eta_{j}}
+\xi_{1}\zeta_{j}S_{22}\partial_{\zeta_{j}}\right)
\right]\psi^{\ell \lambda}_{q1}$$
$$+(q-1)\left[\left(\Omega_{2}S_{j2} - \omega_{j}S_{12}\right)
\partial_{\xi_{j}}
-\omega_{j}S_{22}\partial_{\eta_{j}}
+\zeta_{j}\left(\eta_{2} S_{12}-\xi_{2} S_{22}\right)
\partial_{\zeta_{j}}\right]\psi^{\ell \lambda}_{(q-1)1}$$
$$\left.+(\ell-q)\left[\Omega_{j}S_{11}\partial_{\xi_{j}}
+\left(\Omega_{2} S_{j1}+\Omega_{j} S_{12}\right)
\partial_{\eta_{j}}
+\zeta_{j}\left(-\xi_{2} S_{11}+\xi_{1} S_{12}\right)
\partial_{\zeta_{j}}\right]\psi^{\ell \lambda}_{(q+1)1} \right\}$$
$$=-\left(2/\hbar^{2}\right)\left[E-V\right] \psi^{\ell \lambda}_{q1},
 \eqno (48b) $$

\section{PHYSICAL APPLICATION}

In a quantum $N$-body system, any function with angular 
momentum ${\ell}$ and parity $(-1)^{\ell+\lambda}$ can be
expanded with respect to the generalized harmonic polynomials 
$Q_{q}^{\ell \tau}({\bf R}_{1},{\bf R}_{2})$, where the
coefficients, called the generalized radial functions, 
depend only on $(3N-6)$ internal variables. Since  
$Q_{q}^{\ell \tau}({\bf R}_{1},{\bf R}_{2})$ is a homogeneous
polynomial in the components of the Jacobi coordinate vectors
and a solution of the Laplace equation, we have derived the
generalized radial equations easily. That the rotational variables
(the Euler angles) do not involve in either the generalized 
radial functions or the equations will greatly decrease the
amount of calculation in solving the Schr\"{o}dinger equation 
numerically for the $N$-body system. As a first step, we applied
this approach to the calculation of the energy levels of
a helium atom and a positronium ion \cite{duan1,duan2,duan3}. 
In the following we sketch the method and give some more
calculation results. 

Once the generalized radial equations have been derived, one may
choose any other complete set of internal variables to 
simplify the calculation. The generalized radial equations 
for the new variables can easily be obtained by replacement
of variables. In a Coulombic three-body system, such as
a helium atom, we choose the hyperradius $\rho$ and two 
dimensionless $\eta$ and $\zeta$ as the internal variables, 
so as to make the potential a meromorphic function:
$$\rho=\left({\bf R}_{1}^{2}+{\bf R}_{2}^{2}\right)^{1/2},~~~~~~
\eta=\displaystyle {|{\bf r}_{2}-{\bf r}_{3}| \over \rho },
~~~~~~\zeta=\displaystyle {|{\bf r}_{1}-{\bf r}_{2}| \over \rho }+
\displaystyle {|{\bf r}_{1}-{\bf r}_{3}| \over \rho }, \eqno (49) $$

\noindent
where ${\bf r}_{1}$ denotes the position vector of the helium nucleus
and ${\bf r}_{2}$ and ${\bf r}_{3}$ the position vectors of two
electrons. After expanding the wave function as a Taylor series with 
respect to $\eta$ and $\zeta$ up to the order $n$, we obtain an ordinary
differential matrix equation for the coefficients $R(\rho)$. In 
the real calculation, we calculate the propagating matrix $F(\rho)$ 
and its inverse matrix $G(\rho)$ by the Taylor series method instead 
of the function $R(\rho)$ in order to avoid the logarithmic
singularities at $\rho=0$ in the forms of $\rho^{a}({\rm ln} \rho)^{b}$
\cite{bar,foc,mor}:
$$\rho \displaystyle {dR(\rho) \over d\rho}=F(\rho)R(\rho). \eqno (50) $$

\noindent
We are able to obtain the energy levels of a helium atom in the 
different spectra $^{2S+1}L^{e(o)}$ with high accuracy by 
choosing $n=10$ (for the $S$ and $P$ states) or $n=9$ (for 
the $D$ states) due to the fast convergence of the series, 
where $S$ is the total spin of two electrons, $e(o)$ describes 
the parity, and $L=S$, $P$ and $D$ for the angular momentum 
states.  In order to compare our calculation results with
those by the variational methods where the nucleus mass is
usually assumed to be infinite, we also calculate the energy 
level with a large mass ratio $M$ of the nucleus to the 
electron ($M=10^{20}$). Both calculated results are listed 
in Table 1 for comparison. Much fewer terms in the truncated 
series are taken in our calculation than those in the 
hyperspherical harmonic function method \cite{tang} and in 
the variational methods \cite{drak2}.

\begin{center}
\fbox{Table 1}
\end{center}

\section{CONCLUSIONS}

For a quantum $N$-body system we have found a complete set of
independent base functions $Q_{q}^{\ell \tau}({\bf R}_{1},{\bf R}_{2})$
for the given angular momentum and parity. Any function with
angular momentum ${\ell}$ and parity $(-1)^{\ell+\lambda}$ 
in the system can be expanded with respect to the $(2\ell+1)$ 
generalized harmonic polynomials $Q_{q}^{\ell \tau}({\bf R}_{1},{\bf R}_{2})$,
where the combinative coefficients are functions of the $(3N-6)$
internal variables. We have established the generalized radial
equations explicitly. They are simultaneous partial differential
equations in the internal variables. The number of both the functions
and the equations is $(2\ell+1)$ when $N\geq 4$, and it becomes
$(\ell+1)$ or $\ell$ when $N=3$, depending on the parity.
Only a finite number of partial angular momentum states are
involved in constructing the generalized harmonic polynomials
$Q_{q}^{\ell \tau}({\bf R}_{1},{\bf R}_{2})$. That is, the
contributions from the remaining partial angular momentum
states have been incorporated into those from the generalized
radial functions. We have generalized the formulas to the case
with nonorthogonal vectors.

When establishing the body-fixed frame we fix it with two
arbitrarily chosen Jacobi coordinate vectors ${\bf R}_{1}$ and
${\bf R}_{2}$. Those two vectors may be replaced with any other
two coordinate vectors according to the characteristics of the
physical problem under study.

The choice of the complete set of base functions is not unique.
However, the right choices of both the base functions and the
internal variables play a key role in establishing the generalized
radial equations. Those two choices are the main progress of the
present paper in comparison with the previous work of
Wigner \cite{wig} and Eckart \cite{eck}. Once the generalized 
radial equations have been derived, one may choose any other
complete set of the internal variables to simplify the calculation.
The generalized radial equations for the new variables can easily be 
obtained by replacement of variables, just as we did in Sec. VII
for the three-body system.

The two features in this method, that the numbers of both
functions $\psi^{\ell \lambda}_{q\tau}(\xi, \eta, \zeta)$ and
equations are finite, and they depend on only $(3N-6)$ internal
variables, are important for calculating the energy levels and
wave functions in a quantum $N$-body system. In fact, in the
numerical experiments for a quantum three-body system, we
calculated the lowest energy levels of a helium atom in $P$ states
\cite {duan1} and in $D$ states \cite{duan2} with the total spin
one and zero, and some energy levels of a positronium negative ion
\cite{duan3} by a series expansion. Because three rotational 
variables are removed, many fewer terms have to be taken to achieve
the same precision of energy as in other methods to
truncate the series of partial angular momentum states.
As the number of particles in the system increases, we believe
that removing three independent variables related to the global
rotation will greatly decrease the amount of calculation.

\vspace{2mm}
\noindent
{\bf ACKNOWLEDGMENTS}. The authors would like to thank Professor
Hua-Tung Nieh and Professor Wu-Yi Hsiang for drawing their attention
to quantum few-body problems. This work was supported by
the National Natural Science Foundation of China and the Postdoc 
Science Foundation of China.


\vspace{2mm}

\begin{center}

\item[Table 1]  Numerical calculation for the energy levels
of a helium atom in atomic units

\vspace{3mm} {\footnotesize

\begin{tabular}{|c|l|l|l|}\hline
Spectral Term & \multicolumn{2}{c|}{Our Results}
& Variational Calculation \cite{drak1} \\ \cline{2-3}

$^{2S+1}L^{e(o)}$& $M=7296.28$ & $~~~~M=10^{20}$  
& $~~~~~~~~M\sim \infty$\\ \hline

$^{1}S^{e}$&2.9033046 &2.903724377034116&2.9037243770341195\\

$^{3}S^{e}$&2.1749303& 2.1752293777& 2.1752293782\\
$^{1}P^{e}$&0.5801748&0.5802464725&0.5802465$^{\dagger}$\\
$^{3}P^{e}$&0.7105002&0.7103965&0.710499$^{\dagger}$\\
$^{1}P^{o}$&2.1235456&2.1238430778&2.1238430865\\
$^{3}P^{o}$&2.1328807&2.133164187&2.133164191\\
$^{1}D^{o}$&0.5637256&0.5638004&~~\\
$^{3}D^{o}$&0.5592482&0.5593283&~~\\
$^{3}D^{e}$&2.0553230&2.0555871 &2.0556363 \\
$^{1}D^{e}$&2.0553055&2.0555693&2.0556207\\\hline
\end{tabular}
}
\end{center}
$^{\dagger}$ The calculation in Ref. \cite{drak2}

\end{document}